\documentclass[a4paper, nobib,justified]{tufte-handout}

\usepackage{graphicx} 
\setkeys{Gin}{width=\linewidth,totalheight=\textheight,keepaspectratio}
\graphicspath{{graphics/}} 
\usepackage{amsmath}  
\usepackage{booktabs} 
\usepackage{units}    
\usepackage{multicol} 
\usepackage{lipsum}   
\usepackage{fancyvrb} 
\fvset{fontsize=\normalsize}

\usepackage{tikz} 
\usepackage{subfig}
\usepackage{euler}

\usepackage{orcidlink}

\title{State space modeling of RLC ladder circuits}
\author[Stephan Scholz]{Stephan Scholz \orcidlink{0000-0003-1762-5659}}
\date{}

\begin{document}
	
	\definecolor{wong0}{rgb}{0.0,0.44705883,0.69803923}
\definecolor{wong1}{rgb}{0.9019608,0.62352943,0.0}
\definecolor{wong2}{rgb}{0.0,0.61960787,0.4509804}
\definecolor{wong3}{rgb}{0.8,0.4745098,0.654902}
\definecolor{wong4}{rgb}{0.3372549,0.7058824,0.9137255}
\definecolor{wong5}{rgb}{0.8352941,0.36862746,0.0}
\definecolor{wong6}{rgb}{0.9411765,0.89411765,0.25882354}

\definecolor{rwulila}{cmyk}{0.77,0.79,0,0}
\definecolor{rwucyan}{cmyk}{0.88,0,0.11,0}
\definecolor{myblue}{cmyk}{1,0.67,0,0.4}
\definecolor{mywarm}{cmyk}{0,0.29,0.63,0.16}
\definecolor{petrolblue}{cmyk}{0.98,0.06,0.33,0.47}
\definecolor{leafgreen}{cmyk}{0.65,0,1,0}

\definecolor{juliablue}{rgb}{0.251,0.388,0.847}
\definecolor{juliagreen}{rgb}{0.22,0.596,0.149}
\definecolor{juliared}{rgb}{0.796,0.235,0.2}
\definecolor{juliapurple}{rgb}{0.584,0.345,0.698}

\definecolor{plasmacold}{rgb}{0.050383,0.029803,0.527975}
\definecolor{plasmamiddle}{rgb}{0.794549,0.27577,0.473117}
\definecolor{plasmahot}{rgb}{0.940015,0.975158,0.131326}

\definecolor{wong1}{rgb}{0.0, 0.44705883, 0.69803923}
\definecolor{wong2}{rgb}{0.9019608, 0.62352943, 0.0}
\definecolor{wong3}{rgb}{0.0, 0.61960787, 0.4509804}
\definecolor{wong4}{rgb}{0.8, 0.4745098, 0.654902}
\definecolor{wong5}{rgb}{0.3372549, 0.7058824, 0.9137255}
\definecolor{wong6}{rgb}{0.8352941, 0.36862746, 0.0}
\definecolor{wong7}{rgb}{0.9411765, 0.89411765, 0.25882354}

\pgfdeclarehorizontalshading{plasma}{100bp}{
	color(0bp)=(plasmahot);
	color(25bp)=(plasmahot);
	color(50bp)=(plasmamiddle);
	color(75bp)=(plasmacold);
	color(100bp)=(plasmacold)
}
	
	\maketitle
	
	\begin{abstract}
		\noindent
		Large-scale electrical circuits span a wide range of research topics from highly integrated circuits in microelectronics up to power grids for cities and countries. As the complexity increases significantly by each additional component, we find a need to describe large-scale circuits in a manner easy to understand. In this article, we study cascades of simple circuits consisting of resistors, capacitors and inductors, which we call ladder circuits. Such electrical ladders are common in electronics to design filters and they are also used in other disciplines, for example to describe diffusion or wave phenomena. In particular, we derive linear time-invariant state space models for three simple ladder types and we discuss the specific matrix structures of the large-scale second-order differential equations. Furthermore, we exemplify our findings with simulations to unveil the intrinsic dynamical behavior of each ladder type.
	\end{abstract}

\section*{Introduction}

\begin{marginfigure}[50ex]
	\centering
	\begin{tikzpicture}[scale=1,line cap=round,line join=round,x=1.0cm,y=1.0cm]	
		\def\dxoff{7};
		\def\dyoff{2.8};
		
		\draw[line width=1.3, color=gray] (0,0) -- (0.5,0);
		\draw[line width=1.3, color=gray] (1.5,0) -- (3.5,0);
		\draw[line width=1.3, color=gray] (2.7,0) -- (2.7,-0.7);
		\draw[line width=1.3, color=gray] (2.7,-0.9) -- (2.7,-1.6);
		\draw[line width=1.3, color=gray] (0,-1.6) -- (3.5,-1.6);
		
		\draw [line width=1.0] (-0.1, 0) circle [radius = 0.1];
		\draw [line width=1.0] ( 3.6, 0) circle [radius = 0.1];
		\draw [line width=1.0] (-0.1,-1.6) circle [radius = 0.1];
		\draw [line width=1.0] ( 3.6,-1.6) circle [radius = 0.1];
		
		\draw[line width=2, color=blue,rounded corners] (0.5,-0.2) rectangle (1.5,0.2);

		
		\draw[line width=2, color=blue]  (2.4,-0.7) -- (3.0,-0.7);
		\draw[line width=2, color=blue]  (2.4,-0.9) -- (3.0,-0.9);
		
		\node at ( 1., -0.55) {$R$};
		\node at ( 2.1, -0.8) {$C$};

		
		\draw [->, line width=1.3, color=olive]	(-0.1, -0.2) -- (-0.1, -1.4);
		\draw [->, line width=1.3, color=olive]	( 3.6, -0.2) -- ( 3.6, -1.4);
		\node at (-0.3, -0.7) {$u$};
		\node at ( 3.8, -0.7)  {$y$};
		\node at (-0.6, -0.7) {\rotatebox{90}{Input Voltage}};
		\node at (4.1, -0.7) {\rotatebox{90}{Output Voltage}};
		\node[right] at (0.0, -2.2) {(a) RC Circuit};
		
		\draw [line width=1.3, color=wong1] (0,-1.75-\dyoff) -- (0.3,-1.75-\dyoff);		
		\draw [line width=1.3, yscale=1, domain=0:1.5, smooth, variable=\x, color=wong1] plot ({\x+0.3}, {1.4-1.4*exp(-4*\x)-\dyoff-1.75});
		\draw [line width=1.3, yscale=1, domain=0:2, smooth, variable=\x, color=wong1] plot ({\x+1.8}, {1.4*exp(-4*\x)-\dyoff-1.75});
		
		\draw [->, line width=1.3] (-0.1,-1.8-\dyoff) -- (-0.1,-\dyoff);
		\draw [->, line width=1.3] (-0.1,-1.8-\dyoff) -- (4,-1.8-\dyoff);
		
		\node[right] at (3.1, -2.0-\dyoff) {Time};
		\node[right] at (-0.6, -1.0-\dyoff) {\rotatebox{90}{Voltage}};
		
		\draw [color=wong4, line width=1.2,dashed] (0,-1.75-\dyoff) -- (0.3,-1.75-\dyoff) -- (0.3,-0.3-\dyoff)  -- (1.8,-0.3-\dyoff) -- (1.8,-1.75-\dyoff) -- (3.8,-1.75-\dyoff);
		
		\draw [dashed, line width=1.3, color=wong4] (2.5,-\dyoff-0.3) -- (3.0,-\dyoff-0.3) node[anchor=west] {\color{black} Input u};
		\draw [line width=1.3, color=wong1] (2.5,-\dyoff-0.7) -- (3.0,-\dyoff-0.7) node[anchor=west] {\color{black} Output y};
		
		\node[right] at (0.0, -2.6-\dyoff) {(b) Charging and Discharging};
	\end{tikzpicture}
	\caption{RC Circuit with input voltage $u$ and measured voltage $y$ in (a). Example of charging and discharging voltage dynamics in (b).}
	\label{fig:rc_circuit}
\end{marginfigure} 
As our modern world relies strongly on electrical systems, from nano-scale circuits up to large-scale power grids, studying and designing electrical circuits is a corner stone of our technological developments. Aside explicit engineering application, circuits are also studied to understand and teach fundamental concepts like charging behavior or oscillations in physics, electrical engineering and control theory. For example, circuits with resistors and one capacitor or one inductor (we call them RC and RL circuit)  exemplify charging and discharging dynamics as visualized in Fig. \ref{fig:rc_circuit}; while circuits with two (or more) capacitors or inductors show oscillations. In particular, a LC circuit with ideal inductor and capacitor represents a linear, undamped oscillation of current and voltage.
These observations are joined to the fact that capacitors and inductors store electric and magnetic energy and the number of energy storages in a circuit defines the dynamical behavior and accordingly the order of the differential equation describing this dynamics. In other words, a high number of energy storages imply a high order of the describing differential equation.\sidenote{The number of capacitors or inductors does not have to be equal to the order of the differential equation because in some cases these storages can be summarized, e.g. parallel capacitors or serial inductors.} 

This relation, number of energy storages and order of differential equation, is also used to study different physical phenomena like heat conduction in solids: RC circuits are well known to represent an analogy to thermal dynamics\sidenote{See on Wikipedia \url{https://en.wikipedia.org/wiki/Thermal_conductance_and_resistance} Section "Analogies and nomenclature".}. We find this approach in simple models like the one-dimensional heat conduction, see \cite[p. 62]{book:lienhard2020heat} and \cite{article:berger2019on}, and in advanced systems like an approximation of heat conduction in a hotplate , see Figure 2.2 in \cite[p. 19]{book:han2009temperature}.

In electronics research, we find applications of multiple energy storages in form of passive filters, e.g. LC ladders, which are often described in the frequency domain by complex-valued (rational) functions, see the articles \cite{article:chen2020electrical,article:gerbracht2010on}. On one hand, describing the dynamical behavior in the frequency domain is well established in signal processing and control theory for several decades; on the other hand, modern analysis and design methods rather treat and cover time domain tools like state space models, see the articles \cite{article:pates2022passive,article:piotrowska2021time}. Furthermore, we find in the literature enhanced tools like modified nodal analysis (abbr. as MNA)\sidenote{MNA extends the standard nodal analysis of Kirchhoff's voltage and current law. See wikipedia and this website for further information on MNA:\\ \url{https://lpsa.swarthmore.edu/Systems/Electrical/mna/MNA1.html}} to describe the temporal dynamics of large scale circuits via differential-algebraic system of equations (abbr. as DAE), see the book \cite[p. 27 ff.]{book:najm2010circuit} and the articles \cite{article:tischendorf1998topological,article:tischendorf2001mathematical,article:marz2003finding}. We find such a DAE representation in various fields of mathematical modeling, see \cite{online:campbell2008dae}, and scientific computing, e.g. in software toolkits like \textmd{Modelica}, see the documentation \cite[Appendix B]{online:modelica2026docs}, and \textmd{ModelingTookit.jl}, see the articles \cite{article:ma2021modelingtoolkit,article:rackauckas2022composing}. However, this general applicability comes with the cost of less simplicity (in contrast to ordinary differential equations) and the need of enhanced numerical solvers, e.g. backward differentiation formula (abbr. as BDF), see \cite{online:campbell2008dae,online:gear2007backward} and the book \cite[Chapter VI]{book:wanner1996solving2}.  

In our article, we apply standard nodal analysis on three types of RLC ladders to build systems of ordinary differential equation (abbr. as ODE). In this manner, we avoid the (rather complicated) DAE modeling for a certain class of potentially large scale circuits and we present state space models with specific matrix structure. One of our RLC ladder types, see Fig. \ref{fig:cascaded_rlc_circuits} (a), is considered in the article \cite{article:peters2023dynamic}, to study a machine learning method to compute an unknown state space model from known data.\sidenote{Dynamic mode decomposition is used to find an unknown state space model from observed state trajectories.} We point out that our RLC ladder types 1 and 2,  see Fig. \ref{fig:cascaded_rlc_circuits} (a) and (b), are closely related to the (spatially approximated) telegrapher's equation, which describes the voltage and current dynamics in transmission lines, e.g. see \cite{article:veigend2018model,article:shanak2020mathematical}. Furthermore, we find a similar model of ladder type 1 in the study of port-Hamiltonian systems, see the articles \cite{article:polyuga2010structure,article:gugercin2012structure} and the example ``RCL Ladder Network (ODE)'' in the software library \textit{PortHamiltonianBenchmarkSystems.jl} \cite{software:schwerdtner2025port}.

In Section \ref{sec:circuit_modeling}, we apply Kirchhoff's laws to derive the differential equations for each of the three ladder types. These ODE are formulated in a state space formulation in Section \ref{sec:state_space_model}, where we focus on the embedded matrix shapes. For each ladder, we state an example in Section \ref{sec:simulation_examples} and further, we assume two types of input signals to simulate the temporal dynamics. In Section \ref{sec:telegraph_line}, we show how to extend our modeling approach to the telegraph line model. Finally, we conclude our findings and give an outlook with future developments.

	\section{Circuit Modeling}\label{sec:circuit_modeling}

\begin{figure*}[t!]
	\centering
	\begin{tikzpicture}[scale=1,line cap=round,line join=round,x=1.0cm,y=1.0cm]	
		\def\dxoff{7};
		\def\dyoff{3.8};
		
		\fill[color=green!20,opacity=10, rounded corners] (0.1,-1.7) rectangle (4.9,0.3);
		
		\draw[line width=1.3, color=gray] (0,0) -- (0.5,0);
		\draw[line width=1.3, color=gray] (1.5,0) -- (2.0,0);
		\draw[line width=1.3, color=gray] (2.9,0) -- (5.0,0);
		\draw[line width=1.3, color=gray] (4.3,0) -- (4.3,-0.7);
		\draw[line width=1.3, color=gray] (4.3,-0.9) -- (4.3,-1.6);
		\draw[line width=1.3, color=gray] (0,-1.6) -- (5.,-1.6);
		
		\draw [line width=1.0] (-0.1, 0) circle [radius = 0.1];
		\draw [line width=1.0] ( 5.1, 0) circle [radius = 0.1];
		\draw [line width=1.0] (-0.1,-1.6) circle [radius = 0.1];
		\draw [line width=1.0] ( 5.1,-1.6) circle [radius = 0.1];
		
		\draw[line width=2, color=blue,rounded corners] (0.5,-0.2) rectangle (1.5,0.2);

		\draw[line width=2, color=blue] (2.0,0.) arc ( 180:0:0.15);
		\draw[line width=2, color=blue] (2.3,0.) arc ( 180:0:0.15);
		\draw[line width=2, color=blue] (2.6,0.) arc ( 180:0:0.15);
		
		\draw[line width=2, color=blue]  (4.0,-0.7) -- (4.6,-0.7);
		\draw[line width=2, color=blue]  (4.0,-0.9) -- (4.6,-0.9);
		
		\node at ( 1., -0.55) {$r_{n}$};
		\node at ( 2.45, -0.4) {$\ell_{n}$};
		\node at ( 3.7, -0.8) {$c_{n}$};

		\draw [->, line width=1.3, color=magenta] (3.1,0.05) -- (3.9,0.05);
		\draw [->, line width=1.3, color=magenta] (4.4,0.05) -- (4.9,0.05);
		\draw [->, line width=1.3, color=magenta] (4.25,-0.1) -- (4.25,-0.6);
		\node at (3.5, 0.3) {$i_{m,n}$};
		\node at (4.7, 0.3) {$i_{m,n}$};
		\node at (4.0, -0.3) {$i_{s,n}$};
		
		\draw [->, line width=1.3, color=olive]	(-0.1, -0.2) -- (-0.1, -1.4);
		\draw [->, line width=1.3, color=olive]	( 5.1, -0.2) -- ( 5.1, -1.4);
		\draw [->, line width=1.3, color=olive]	( 0.45, 0.5) -- ( 2.95, 0.5);
		\node at (-0.5, -0.7) {$z_{n-1}$};
		\node at ( 5.5, -0.7)  {$z_{n}$};
		\node at ( 1.6, 0.7) {$v_{n}$};
		
		\node[anchor=west] at (0.3, -2.2) {(a) Type 1: Capacitor as Shunt};

		\draw[line width=1.3, color=gray] (\dxoff+0,0) -- (\dxoff+0.8,0);
		\draw[line width=1.3, color=gray] (\dxoff+1.82,0) -- (\dxoff+2.7,0);
		\draw[line width=1.3, color=gray] (\dxoff+2.9,0) -- (\dxoff+5.0,0);
		\draw[line width=1.3, color=gray] (\dxoff+4.3,0) -- (\dxoff+4.3,-0.3);
		\draw[line width=1.3, color=gray] (\dxoff+4.3,-1.3) -- (\dxoff+4.3,-1.6);
		\draw[line width=1.3, color=gray] (\dxoff+0,-1.6) -- (\dxoff+5.0,-1.6);

		\draw [line width=1.0] (\dxoff-0.1, 0) circle [radius = 0.1];
		\draw [line width=1.0] (\dxoff+5.1, 0) circle [radius = 0.1];
		\draw [line width=1.0] (\dxoff-0.1,-1.6) circle [radius = 0.1];
		\draw [line width=1.0] (\dxoff+5.1,-1.6) circle [radius = 0.1];
		
		\draw [->, line width=1.3, color=olive]	(\dxoff -0.1, -0.2) -- (\dxoff -0.1, -1.4);
		\draw [->, line width=1.3, color=olive]	(\dxoff +5.1, -0.2) -- (\dxoff +5.1, -1.4);
		\draw [->, line width=1.3, color=olive]	(\dxoff +0.45, 0.5) -- (\dxoff +3.1, 0.5);
		
		\node at (\dxoff -0.5, -0.7) {$z_{n-1}$};
		\node at (\dxoff+ 5.5, -0.7)  {$z_{n}$};
		\node at (\dxoff+ 1.6, 0.7) {$v_{n}$};
		
		\draw[line width=2, color=blue] (\dxoff+0.8, 0) arc ( 180:0:0.17);
		\draw[line width=2, color=blue] (\dxoff+1.14, 0) arc ( 180:0:0.17);
		\draw[line width=2, color=blue] (\dxoff+1.48, 0) arc ( 180:0:0.17);
		
		\draw[line width=2, color=blue] (\dxoff+2.7,-0.2) -- (\dxoff+2.7,0.2);
		\draw[line width=2, color=blue] (\dxoff+2.9,-0.2) -- (\dxoff+2.9,0.2);
		
		\draw[line width=2, color=blue,rounded corners] (\dxoff+4.1,-0.3) rectangle (\dxoff+4.5,-1.3);
		
		\node at (\dxoff+ 1.4, -0.5) {$\ell_{n}$};
		\node at (\dxoff+ 2.8, -0.5) {$c_{n}$};
		\node at (\dxoff+ 3.7, -0.9) {$r_{n}$};
		
		\node[anchor=west] at (\dxoff+0.3, -2.2) {(b) Type 2: Resistor as Shunt};
		
		
		\draw[line width=1.3, color=gray] (  0, 0-\dyoff) -- (0.8,0-\dyoff);
		\draw[line width=1.3, color=gray] (1.8, 0-\dyoff) -- (2.7,0-\dyoff);
		\draw[line width=1.3, color=gray] (2.9, 0-\dyoff) -- (5.0,0-\dyoff);
		\draw[line width=1.3, color=gray] (4.3, 0-\dyoff) -- (4.3,-0.3-\dyoff);
		\draw[line width=1.3, color=gray] (4.3,-1.2-\dyoff) -- (4.3,-1.6-\dyoff);
		
		\draw[line width=1.3, color=gray] (  0,-1.6-\dyoff) -- (5.0,-1.6-\dyoff);
		
		\draw [line width=1.0] (-0.1, 0-\dyoff) circle [radius = 0.1];
		\draw [line width=1.0] ( 5.1, 0-\dyoff) circle [radius = 0.1];
		\draw [line width=1.0] (-0.1,-1.6-\dyoff) circle [radius = 0.1];
		\draw [line width=1.0] ( 5.1,-1.6-\dyoff) circle [radius = 0.1];
		
		\draw [->, line width=1.3, color=olive]	(-0.1, -0.2-\dyoff) -- (-0.1, -1.4-\dyoff);
		\draw [->, line width=1.3, color=olive]	( 5.1, -0.2-\dyoff) -- ( 5.1, -1.4-\dyoff);
		\draw [->, line width=1.3, color=olive]	( 0.45, 0.5-\dyoff) -- ( 3.1, 0.5-\dyoff);
		
		\node at (-0.5, -0.7-\dyoff) {$z_{n-1}$};
		\node at ( 5.5, -0.7-\dyoff)  {$z_{n}$};
		\node at ( 1.6, 0.7-\dyoff) {$v_{n}$};
		
		\draw[line width=2, color=blue,rounded corners] (0.8,-0.2-\dyoff) rectangle (1.8,0.2-\dyoff);
		
		\draw[line width=2, color=blue] (2.7,-0.2-\dyoff) -- (2.7,0.2-\dyoff);
		\draw[line width=2, color=blue] (2.9,-0.2-\dyoff) -- (2.9,0.2-\dyoff);
		
		\draw[line width=2, color=blue] (4.3,-0.3-\dyoff) arc ( 90:-90:0.15);
		\draw[line width=2, color=blue] (4.3,-0.6-\dyoff) arc ( 90:-90:0.15);
		\draw[line width=2, color=blue] (4.3,-0.9-\dyoff) arc ( 90:-90:0.15);

		\node at ( 1.4, -0.5-\dyoff) {$r_{n}$};
		\node at ( 2.8, -0.5-\dyoff) {$c_{n}$};
		\node at ( 3.8, -0.9-\dyoff) {$\ell_{n}$};
		
		\node[anchor=west] at (0.3, -2.3-\dyoff) {(c) Type 3: Inductor as Shunt};

		
		\draw[line width=1.3, color=gray] (\dxoff+0  , 0-\dyoff) -- (\dxoff+0.3, 0-\dyoff);
		\draw[line width=1.3, color=gray] (\dxoff+0  ,-1.6-\dyoff) -- (\dxoff+0.3,-1.6-\dyoff);
		\draw[line width=1.3, color=gray] (\dxoff+1.2, 0-\dyoff) -- (\dxoff+1.6, 0-\dyoff);
		\draw[line width=1.3, color=gray] (\dxoff+1.2,-1.6-\dyoff) -- (\dxoff+1.6,-1.6-\dyoff);
		\draw[line width=1.3, color=gray] (\dxoff+1.8, 0-\dyoff) -- (\dxoff+2.3, 0-\dyoff);
		\draw[line width=1.3, color=gray] (\dxoff+1.8,-1.6-\dyoff) -- (\dxoff+2.3,-1.6-\dyoff);
		\draw[line width=1.3, color=gray] (\dxoff+3.2, 0-\dyoff) -- (\dxoff+3.6, 0-\dyoff);
		\draw[line width=1.3, color=gray] (\dxoff+3.2,-1.6-\dyoff) -- (\dxoff+3.6,-1.6-\dyoff);			
		\draw[line width=1.3, color=gray] (\dxoff+3.8, 0-\dyoff) -- (\dxoff+4.1, 0-\dyoff);
		\draw[line width=1.3, color=gray] (\dxoff+3.8,-1.6-\dyoff) -- (\dxoff+4.1,-1.6-\dyoff);			
		\draw[line width=1.3, color=gray] (\dxoff+4.3, 0-\dyoff) -- (\dxoff+4.6, 0-\dyoff);
		\draw[line width=1.3, color=gray] (\dxoff+4.3,-1.6-\dyoff) -- (\dxoff+4.6,-1.6-\dyoff);
		\draw[line width=1.3, color=gray] (\dxoff+4.8, 0-\dyoff) -- (\dxoff+5.8, 0-\dyoff);
		\draw[line width=1.3, color=gray] (\dxoff+4.8,-1.6-\dyoff) -- (\dxoff+5.8,-1.6-\dyoff);
		\draw[line width=1.3, color=gray] (\dxoff+6.4, 0-\dyoff) -- (\dxoff+7, 0-\dyoff);
		\draw[line width=1.3, color=gray] (\dxoff+6.4,-1.6-\dyoff) -- (\dxoff+7,-1.6-\dyoff);	
		
		\draw [line width=1.0] (\dxoff+-0.1, 0-\dyoff) circle [radius = 0.1];
		\draw [line width=1.0] (\dxoff+-0.1,-1.6-\dyoff) circle [radius = 0.1];
		
		\draw [line width=1.0] (\dxoff+1.7, 0-\dyoff) circle [radius = 0.1];
		\draw [line width=1.0] (\dxoff+ 1.7,-1.6-\dyoff) circle [radius = 0.1];
		
		\draw [line width=1.0] (\dxoff+3.7, 0-\dyoff) circle [radius = 0.1];
		\draw [line width=1.0] (\dxoff+ 3.7,-1.6-\dyoff) circle [radius = 0.1];
		
		\draw [line width=1.0] (\dxoff+ 4.7, 0-\dyoff) circle [radius = 0.1];
		\draw [line width=1.0] (\dxoff+ 4.7,-1.6-\dyoff) circle [radius = 0.1];
		
		\draw [line width=1.0] (\dxoff+ 6.8+0.3, 0-\dyoff) circle [radius = 0.1];
		\draw [line width=1.0] (\dxoff+ 6.8+0.3,-1.6-\dyoff) circle [radius = 0.1];
		
		\draw [line width=1.0] (\dxoff+4,   -0.2-\dyoff) -- (\dxoff+4.25, 0.2-\dyoff);	
		\draw [line width=1.0] (\dxoff+4.15,-0.2-\dyoff) -- (\dxoff+4.4 , 0.2-\dyoff);	
		\draw [line width=1.0] (\dxoff+4,   -1.8-\dyoff) -- (\dxoff+4.25,-1.4-\dyoff);	
		\draw [line width=1.0] (\dxoff+4.15,-1.8-\dyoff) -- (\dxoff+4.4 ,-1.4-\dyoff);		
		
		\draw[dashed,line width=2,rounded corners, fill=green!20] (\dxoff+0.3,-1.9-\dyoff) rectangle (\dxoff+1.2,0.3-\dyoff);
		\draw[dashed,line width=2,rounded corners, fill=green!20] (\dxoff+2.3,-1.9-\dyoff) rectangle (\dxoff+3.2,0.3-\dyoff);
		\draw[dashed,line width=2,rounded corners, fill=green!20] (\dxoff+5.5+0.3,-1.9-\dyoff) rectangle (\dxoff+6.4+0.3,0.3-\dyoff);
		
		\node at (\dxoff+0.77,-0.6-\dyoff) {RLC};
		\node at (\dxoff+0.77,-1.0-\dyoff) {\textbf{1}};
		
		\node at (\dxoff+2.77,-0.6-\dyoff) {RLC};
		\node at (\dxoff+2.77,-1.0-\dyoff) {\textbf{2}};
		
		\node at (\dxoff+5.97+0.3,-0.6-\dyoff) {RLC};
		\node at (\dxoff+5.97+0.3,-1.0-\dyoff) {\textbf{N}};
		
		\draw [->, line width=1.3, color=olive]	(\dxoff+-0.1, -0.2-\dyoff) -- (\dxoff+-0.1, -1.4-\dyoff);
		\draw [->, line width=1.3, color=olive]	(\dxoff+ 1.7, -0.2-\dyoff) -- (\dxoff+ 1.7, -1.4-\dyoff);
		\draw [->, line width=1.3, color=olive]	(\dxoff+ 3.7, -0.2-\dyoff) -- (\dxoff+ 3.7, -1.4-\dyoff);
		\draw [->, line width=1.3, color=olive]	(\dxoff+ 4.7, -0.2-\dyoff) -- (\dxoff+ 4.7, -1.4-\dyoff);
		\draw [->, line width=1.3, color=olive]	(\dxoff+ 6.8+0.3, -0.2-\dyoff) -- (\dxoff+ 6.8+0.3, -1.4-\dyoff);
		\node at (\dxoff+-0.3, -0.8-\dyoff) {$u$};
		\node at (\dxoff+ 1.95, -0.8-\dyoff) {$z_{1}$};
		\node at (\dxoff+ 3.95, -0.8-\dyoff) {$z_{2}$};
		\node at (\dxoff+ 5.2, -0.8-\dyoff) {$z_{N-1}$};
		\node at (\dxoff+ 7.5+0.3, -0.8-\dyoff)  {$y=z_{N}$};
		
		\node[anchor=west] at (\dxoff+0.3, -2.3-\dyoff) {(d) Circuit Ladder};
		
	\end{tikzpicture}
	\caption{Considered types of RLC circuits (a-c) and a ladder of such circuits (d). The voltage at the last shunt is the output, $z_{n} = y$.}
	\label{fig:cascaded_rlc_circuits}
	
\end{figure*}

\begin{table}[b!]
	\centering
	\renewcommand{\arraystretch}{1.3}
	\fontfamily{ppl}\selectfont
	\begin{tabular}{lll}
		\toprule
		\textbf{Sym.} & \textbf{Explanation}  & \textbf{Domain}  \\
		\midrule
		$u$ & Input voltage & $[0,T] \rightarrow \mathbb{R}$ \\
		$y$ & Output voltage & $[0,T] \rightarrow \mathbb{R}$ \\
		$v_{n}$ & Voltage at main branch & $[0,T] \rightarrow \mathbb{R}$ \\
		$z_{n}$ & Voltage at shunt & $[0,T] \rightarrow \mathbb{R}$ \\
		$i_{m,n}$ & Current through main branch & $[0,T] \rightarrow \mathbb{R}$ \\
		$i_{s,n}$ & Current through shunt & $[0,T] \rightarrow \mathbb{R}$ \\
		$r_{n}$, $\ell_{n}$, $c_{n}$ & Resistor, Inductor, Capacitor & $\mathbb{R}_{>0} \setminus \{\infty\}$ \\
		\bottomrule
	\end{tabular}
	\caption{Nomenclature of electrical states and quantities.}
	\label{table:el_symbols_explanation}
\end{table}

We begin the modeling with a single circuit of a ladder and demonstrate afterwards how to scale this approach for a specific RLC ladder. In our modeling approach, we consider nominal electronic components, i.e. a resistor has only a pure resistance without inductance or capacitance; and we assume a voltage source, which can supply an arbitrary signal. The voltages, currents and the quantities of the electrical components are noted in Table \ref{table:el_symbols_explanation}.\sidenote{We emphasize that we note the electrical coefficients in lower case letters here: $r$, $\ell$, $c$ instead of $R$, $L$, $C$.} The currents and voltages are functions of time, e.g. $u: [0,T] \rightarrow\mathbb{R}$ with final time $T>0$, while the quantities 	$r_{n}$, $\ell_{n}$ and $c_{n}$ are constant. We summarize the voltages at the shunt as 
\begin{equation}
	z = \left[z_{1}, z_{2}, \ldots, z_{N}\right]^{\top} \label{eq:z_voltages_shunt}
\end{equation}	
and we consider the voltage at the last shunt as the output voltage $y(t) = z_{N}(t)$. We consider the currents and voltages to be sufficiently smooth, e.g. twice smoothly differentiable $i_{n}, v_{n}, z_{n} \in \mathcal{C}^{2}([0,T])$, and integrable, e.g. $\int_{0}^{T} i_{n}(\tau) d\tau < \infty$. According to Kirchhoff's voltage law, the $n$-th shunt voltage is identified by
\begin{equation}
	z_{n-1}(t) = v_{n}(t) + z_{n}(t) = \left[\sum_{k=n}^{N} v_{k}(t)\right] + z_{N}(t) \label{eq:kirchhoff_voltage}
\end{equation}
and the input voltage in the first circuit is $z_{0} = u(t)$. As a ladder is a cascade of parallel circuits, the currents split up to the subsequent circuit $n+1$ and the shunt as
\begin{equation}
	i_{m,n}(t) =~ i_{m,n+1}(t) + i_{s,n}(t) ~=~ \sum_{k=n}^{N} i_{s,k}(t) \label{eq:kirchhoff_current}
\end{equation}
with the current in the final circuit $i_{m,N}(t) = i_{s,N}(t)$. We know that the ideal voltage $x:[0,\infty) \rightarrow \mathbb{R}$ across the electrical components is stated as 
\begin{align*}
	\text{Resistor:} \quad x(t) =&~ r ~ i(t) ~\text{,}\\
	\text{Inductor:} \quad x(t) =&~ \ell ~ \partial_{t} i(t) ~\text{,}\\
	\text{Capacitor:} \quad x(t) =&~ \frac{1}{c} ~ \int_{0}^{t} i(\tau) d\tau \text{.}
\end{align*}
These identities and compositions according to Eq. \eqref{eq:kirchhoff_voltage} are modeled as linear operator mappings 
\begin{equation*}
	\mathcal{F}\{i\}(t) = a_{0} ~ i(t) +  a_{1}~ \partial_{t} i(t) +  a_{2}~ \int_{0}^{t} i(\tau) d\tau 
\end{equation*}
with coeffients $a_{0},~a_{1},~a_{2} \in \mathbb{R}_{\geq 0}$ corresponding to our electrical quantities (or its inverse); and we distinguish these mappings for the main branch and the shunt as
\begin{equation}
	v_{n}(t) = \mathcal{F}_{m,n} \left\{i_{m,n}\right\}(t) ~ \text{,} \quad 
	z_{n}(t) = \mathcal{F}_{s,n} \left\{i_{s,n}\right\}(t) \text{.} \label{eq:operator_maps}
\end{equation}
Subsequently, we apply the Kirchhoff laws and the operator mappings on the considered types of RLC ladders, see Fig. \ref{fig:cascaded_rlc_circuits}, in order to derive the systems of linear ordinary differential equations.

\subsection{Ladder Type 1: Capacitor as Shunt}

In the first case, see Fig. \ref{fig:cascaded_rlc_circuits} (a), the voltage in the main branch with a resistor and an inductor is stated as
\begin{align}
	v_{n}(t) =&~ r_n i_{m,n}(t) + \ell_n \partial_{t} i_{m,n}(t) \nonumber \\ 
	=&~ (r_{n} + \ell_{n}\partial_{t}) ~ i_{m,n}(t) = \mathcal{F}_{m,n} \left\{i_{m,n}\right\}(t) \nonumber \\ 
	=&~ \left[r_{n} + \ell_{n}\partial_{t}\right] ~ \sum_{k=n}^{N} i_{s,k}(t) \text{,} \label{eq:rlc_cap_volt_r_l}
\end{align}
and the shunt voltage is noted as
\begin{equation}
	z_{n}(t) =~ \frac{1}{c_{n}} \int_{0}^{t} i_{s,n}(\tau) d\tau ~~= \mathcal{F}_{s,n} \left\{i_{s,n}\right\}(t) \text{.} \label{eq:rlc_cap_volt_c}
\end{equation}
We compute the inverse operator mapping of Eq. \eqref{eq:rlc_cap_volt_c} as 
\begin{equation*}
	i_{s,n}(t) = \mathcal{F}_{s,n}^{-1} \left\{z_{n}\right\}(t) = c_{n} \partial_{t} z_{n}(t)	
\end{equation*}
and insert it in Eq. \eqref{eq:rlc_cap_volt_r_l} to yield
\begin{align}
	v_{n}(t) =&~  \left[r_{n} + \ell_{n}\partial_{t}\right] ~ \sum_{k=n}^{N} c_{k} ~ \partial_{t} z_{k}(t) \nonumber \\
	=&~  \sum_{k=n}^{N} \left[ r_{n} ~  c_{k} ~  \partial_{t} + \ell_{n} ~  c_{k} ~ \partial^2_{t}\right] z_{k}(t) \label{eq:rlc_cap_volt_r_l_2}
\end{align}
We consider Kirchhoff's voltage law \eqref{eq:kirchhoff_voltage} with the voltage in main branch \eqref{eq:rlc_cap_volt_r_l_2} and so we calculate the shunt voltage for the $n$-th circuit, $n\in\{2,\ldots,N\}$ as the differential equation
\begin{align}
	0 =&~ \sum_{k=n}^{N} \left[ \ell_{n} c_{k} ~ \partial_{t}^2 z_{k}(t) + r_{n} c_{k} ~ \partial_{t} z_{k}(t)  \right] \nonumber \\
	&\qquad  + z_{n}(t) - z_{n-1}(t) \text{.} \label{eq:rl-c_shunt_ode_n}
\end{align}
For the first circuit of the ladder, we have $u(t) = z_{0}(t)$ and so we note
\begin{equation}
	u(t) =~ \sum_{k=1}^{N} \left[ \ell_{1} c_{k} ~ \partial_{t}^2 z_{k}(t) + r_{1} c_{k} ~ \partial_{t} z_{k}(t)  \right] + z_{1}(t) \text{.} \label{eq:rl-c_shunt_ode_0}
\end{equation}

\subsection{Ladder Type 2: Resistor as Shunt}
In the second case, see Fig. \ref{fig:cascaded_rlc_circuits} (b), we have an inductor and a capacitor in the main branch and a resistor as shunt. Hence, we note the voltages as 
\begin{align} 
	v_{n}(t) =&~ \ell_n ~ \partial_{t} i_{m,n}(t) + \frac{1}{c_{n}} \int_{0}^{t} i_{m,n}(\tau) d\tau \nonumber \\
	=&~ \mathcal{F}_{m,n} \left\{i_{m,n}\right\}(t) \qquad\text{and} \label{eq:lc-r_volt_main} \\
	z_{n}(t) =&~ r_n ~ i_{s,n}(t)  = \mathcal{F}_{s,n} \left\{i_{s,n}\right\}(t) \text{.} \label{eq:lc-r_volt_shunt}
\end{align}
We differentiate both sides of Eq. \eqref{eq:lc-r_volt_main} and we apply Eq. (\ref{eq:kirchhoff_current}, \eqref{eq:lc-r_volt_shunt}) to obtain
\begin{align*}
	\partial_{t} v_{n}(t) =&~ \left[\ell_n \partial_{t}^2 + \frac{1}{c_{n}}\right] i_{m,n}(t) \\
	=&~ \left[\ell_n \partial_{t}^2 + \frac{1}{c_{n}}\right] \left[  \sum_{k=n}^{N} i_{s,k}(t)  \right] \\
	=&~ \left[\ell_n \partial_{t}^2 + \frac{1}{c_{n}}\right] \left[  \sum_{k=n}^{N} \frac{z_{k}(t)}{r_{k}} \right] \text{.}
\end{align*}
We evaluate the Kirchhoff voltage law \eqref{eq:kirchhoff_voltage} and we  note the differential equation of the $n$-th circuit 
for  $k\in\{2,\ldots,N\}$ as 
\begin{align}
	0 =&~ \sum_{k=n}^{N} \left[ \frac{\ell_{n}}{r_{k}} ~ \partial_{t}^2 z_{k}(t) + \frac{z_{k}(t)}{r_{k}~c_{n}} \right] \nonumber \\
	&\qquad   + \partial_{t} \left[z_{n}(t) - z_{n-1}(t)\right] \label{eq:lc-r_shunt_ode_n}
\end{align}
for $n\in\{2,\ldots,N\}$, and for $n=1$ we have
\begin{equation}
	\partial_{t} u(t) =~ \sum_{k=1}^{N} \left[ \frac{\ell_{1}}{r_{k}} ~ \partial_{t}^2 z_{k}(t) + \frac{z_{k}(t)}{r_{k}~c_{1}} \right] + \partial_{t} z_{1}(t) \text{.} \label{eq:lc-r_shunt_ode_0}
\end{equation}

\subsection{Ladder Type 3: Inductor as Shunt}
In case of the third circuit type, see Fig. \ref{fig:cascaded_rlc_circuits} (c), we note the voltages as 
\begin{align}
	v_{n}(t) =&~ r_n ~ i_{m,n}(t) + \frac{1}{c_{n}} \int_{0}^{t} i_{m,n}(\tau) d\tau \text{,} \label{eq:rc-l_volt_main}  \\ 
	z_{n}(t) =&~ \ell_n ~ \partial_{t} i_{s,n}(t) \text{.} \label{eq:rc-l_volt_shunt}
\end{align}
We differentiate Eq. \eqref{eq:rc-l_volt_main} two times and we insert $\partial_{t} i_{s,n}(t)$ from Eq. \eqref{eq:rc-l_volt_shunt}. Hence, we yield the identity
\begin{align*}
	\partial_{t}^2 v_{n}(t) =&~ \left[r_n ~ \partial_{t} + \frac{1}{c_{n}} \right] \partial_{t}  i_{m,n}(t) \\
	=&~ \left[r_n ~ \partial_{t} + \frac{1}{c_{n}} \right]  \left[ \sum_{k=n}^{N} \frac{z_{k}(t)}{\ell_{k}}  \right] \text{.}
\end{align*}
In a analog way as before, we note the differential equation of the $n-th$ circuit as
\begin{align}
	0 =&~  \partial_{t}^2 \left[z_{n}(t) - z_{n-1}(t)\right] \nonumber \\
	&\quad + \sum_{k=n}^{N} \left[ \frac{r_{n}}{\ell_{k}}~\partial_{t} z_{k}(t) + \frac{z_{k}(t)}{\ell_{k}~c_{n}} \right] \label{eq:rc-l_shunt_ode_n}
\end{align}
and for the first circuit, we note
\begin{equation}
	\partial_{t}^2 u(t) =~  \partial_{t}^2 z_{1}(t) + \sum_{k=n}^{N} \left[ \frac{r_{1}}{\ell_{k}}~\partial_{t} z_{k}(t) + \frac{z_{k}(t)}{\ell_{k}~c_{1}} \right] \text{.} \label{eq:rc-l_shunt_ode_0}
\end{equation}
	
\section{State Space Model}
\label{sec:state_space_model}

We arrange our ODE from the Kirchhoff laws in the previous section to form a second-order differential equation with the two noticable matrix shapes and afterwards, we calculate the system matrix of the state-space model for each ladder type in a closed form.

\subsection{Second-Order System}
\label{sec:state_space_model_second_order}

We collect the equations of the shunt voltages, e.g. Eq.  (\ref{eq:rl-c_shunt_ode_0}, \ref{eq:rl-c_shunt_ode_n}), and we arrange them as a system of second-order equations
\begin{equation}
	\mathbf{M}~ \ddot{z}(t) + \mathbf{D}~ \dot{z}(t) + \mathbf{S}~ z(t) = \mathbf{G} ~ \mu(t)
	\label{eq:rlc_second_order_1}
\end{equation}
with matrices $\mathbf{M},~\mathbf{D},~\mathbf{S} \in \mathbb{R}^{N\times N}$, and vectors $z$ as in Eq. \eqref{eq:z_voltages_shunt} and  $\mathbf{G} = [1, 0, \ldots, 0]^{\top}$. The indices of our matrix elements denote a row $n\in \{1,\ldots,N\}$ and column $k\in \{1,\ldots,N\}$. This index notation refers to the indices in the voltage differential equations, e.g. Eq. (\ref{eq:rl-c_shunt_ode_n}, \ref{eq:rl-c_shunt_ode_0}) etc. In Eq. \eqref{eq:rlc_second_order_1}, we need to distinguish the input signal for each ladder type as
\begin{equation}
	\mu(t) = 
	\begin{cases}
		~~~u(t) \quad &\text{for ladder type 1, see Eq. \eqref{eq:rl-c_shunt_ode_0},} \\
		\partial_{t} u(t) \quad &\text{for ladder type 2, see Eq. \eqref{eq:lc-r_shunt_ode_0} and}\\
		\partial_{t}^2 u(t) \quad &\text{for ladder type 3, see Eq. \eqref{eq:rc-l_shunt_ode_0}.}\\
	\end{cases}
	\label{eq:input_signal}
\end{equation}

The symbols of these matrices, $\mathbf{M}$, $\mathbf{D}$ and $\mathbf{S}$, refer to the mass, damping and stiffness in mechanical systems. Our modeling approach of the second-order systems results in two types of matrices: upper triangular shapes and lower finite difference stencils, see the matrix shapes in Fig. \ref{fig:matrix_shapes}. The upper triangular matrices have a shape as
\begin{equation}
	\mathbf{T} = 
	\begin{pmatrix}
		a_{1} b_{1} & a_{1} b_{2} & \cdots &  a_{1} b_{N-1} & a_{1} b_{N} \\
		0 			& a_{2} b_{2} & \cdots &  a_{2} b_{N-1} & a_{2} b_{N} \\
		\vdots			& \ddots	  & \ddots &  \vdots        & 	\vdots	  \\
		&			  &        & a_{N-1} b_{N-1}& a_{N-1} b_{N} \\
		0			&	\cdots		  &        &	0            & a_{N} b_{N} 
	\end{pmatrix}
	\label{eq:triangular_matrix} 
\end{equation}
where our coefficients $(a_{n}, b_{k})$ with $n,k \in \{1,\ldots,N\}$ stand for the related electrical coefficients and their reciprocals, see Equations (\ref{eq:rl-c_shunt_ode_n}, \ref{eq:rl-c_shunt_ode_0}), (\ref{eq:lc-r_shunt_ode_n}, \ref{eq:lc-r_shunt_ode_0}) and (\ref{eq:rc-l_shunt_ode_n},\ref{eq:rc-l_shunt_ode_0}), and Table \ref{table:triangular_matrix_values}. The finite difference stencil is
\begin{equation}
	\mathbf{F} = 
	\begin{pmatrix}
		1 & 0 & \cdots &  &  0\\
		-1 			& 1 & 0 &   & \vdots  \\
		0 & \ddots	  & \ddots & \ddots  & 	 \\
		\vdots &			  &     -1  & 1 &  0 \\
		0 &	\cdots		  &     0   & -1  & 1 
	\end{pmatrix} 
	\textbf{.}
	\label{eq:finite_diff_stencil} 
\end{equation}
The matrix entries for each circuit type are noted in Table \ref{table:triangular_matrix_values}. The triangular matrix structure corresponds to a spatially discrete integration and the finite difference shape to a differentiation.  We assume, $r_{n}$, $\ell_{n}$, $c_{n} \in \mathbb{R}_{>0} \setminus \{\infty\}$ and so the resulting triangular matrices are invertible. Hence, we are able to reformulate the second-order system via inverse matrix multiplication as first-order system. 

\begin{margintable}
	\renewcommand{\arraystretch}{1.3}
	\centering
	\caption{{Triangular Matrix Corresponding Values}}
	\begin{tabular}{l l l l}
		\toprule
		Shunt   & Matrix & $a_{n}$ & $b_{k}$ \\
		\midrule
		Capacitor   & $\mathbf{M}$ & $\ell_{n}$   & $c_{k}$	\\
		~~ (Type 1) & $\mathbf{D}$ & $r_{n}$      & $c_{k}$ \\[1.5ex] 
		Resistor    & $\mathbf{M}$ & $\ell_{n}$   & $r_{k}^{-1}$ \\ 
		~~ (Type 2) & $\mathbf{S}$ & $c_{n}^{-1}$ & $r_{k}^{-1}$ \\[1.5ex] 
		Inductor 	& $\mathbf{D}$ & $r_{n}$      & $\ell_{k}^{-1}$ \\
		~~ (Type 3) & $\mathbf{S}$ & $c_{n}^{-1}$ & $\ell_{k}^{-1}$ \\ 
		\bottomrule
	\end{tabular}
	\label{table:triangular_matrix_values}	
\end{margintable}

\begin{figure}[t!]
	\centering
	\begin{tikzpicture}[scale=1,line cap=round,line join=round,x=1.0cm,y=1.0cm]	
		\def\dx{2};
		\def\dy{2.2};
		
		\foreach \i in {0, 1, 2}
		{
			\draw[line width=1.0] (0.1+\i*\dx,0+\dy) -- (0+\i*\dx,0+\dy) -- (0+\i*\dx,1+\dy) -- (0.1+\i*\dx,1+\dy);
			\draw[line width=1.0] (1.0+\i*\dx,0+\dy) -- (1.1+\i*\dx,0+\dy) -- (1.1+\i*\dx,1+\dy) -- (1.0+\i*\dx,1+\dy);
		}
		\node at (1.5,0.5+\dy) {\Large $\mathbf{\ddot{z}}+$};
		\node at (1.5+\dx,0.5+\dy) {\Large $\mathbf{\dot{z}}+$};
		\node at (1.55+2*\dx,0.47+\dy) {\Large $\mathbf{z}=$};
		
		\draw[line width=1, color=blue, fill=gray!10] (0.1, 0.9+\dy) -- (1.0, 0.9+\dy) -- (1.0, 0.1+\dy) -- cycle;
		\draw[line width=1, color=blue, fill=gray!10] (0.1+\dx, 0.9+\dy) -- (1.0+\dx, 0.9+\dy) -- (1.0+\dx, 0.1+\dy) -- cycle;
		\draw[dash dot,line width=1.5, color=blue] (0.1+2*\dx,0.9+\dy) -- (1.+2*\dx,0.1+\dy);
		\draw[dashed,line width=1.5, color=orange] (0.1+2*\dx,0.7+\dy) -- (0.8+2*\dx,0.1+\dy);
		\draw[line width=1.0] (0.2+3*\dx,0+\dy) -- (0.1+3*\dx,0+\dy) -- (0.1+3*\dx,1+\dy) -- (0.2+3*\dx,1+\dy);
		\draw[line width=1.0] (0.3+3*\dx,0+\dy) -- (0.4+3*\dx,0+\dy) -- (0.4+3*\dx,1+\dy) -- (0.3+3*\dx,1+\dy);
		\node at (0.25+3*\dx,0.8+\dy) {\large $1$};
		\node at (0.6+3*\dx,0.52+\dy) {\Large $\mathbf{u}$};
		
		\node[right] at (1.0, -0.5+\dy) {(a) Capacitor as shunt};

		\foreach \i in {0, 1, 2}
		{
			\draw[line width=1.0] (0.1+\i*\dx,0) -- (0+\i*\dx,0) -- (0+\i*\dx,1) -- (0.1+\i*\dx,1);
			\draw[line width=1.0] (1.0+\i*\dx,0) -- (1.1+\i*\dx,0) -- (1.1+\i*\dx,1) -- (1.0+\i*\dx,1);
		}
		\node at (1.5,0.5) {\Large $\mathbf{\ddot{z}}+$};
		\node at (1.5+\dx,0.5) {\Large $\mathbf{\dot{z}}+$};
		\node at (1.55+2*\dx,0.47) {\Large $\mathbf{z}=$};
		\draw[line width=1, color=blue, fill=gray!10] (0.1, 0.9) -- (1.0, 0.9) -- (1.0, 0.1) -- cycle;
		\draw[line width=1, color=blue, fill=gray!10] (0.1+2*\dx, 0.9) -- (1.0+2*\dx, 0.9) -- (1.+2*\dx, 0.1) -- cycle;
		\draw[dash dot,line width=1.5, color=blue] (0.1+\dx,0.9) -- (1.+\dx,0.1);
		\draw[dashed,line width=1.5, color=orange] (0.1+\dx,0.7) -- (0.8+\dx,0.1);
		\draw[line width=1.0] (0.2+3*\dx,0) -- (0.1+3*\dx,0) -- (0.1+3*\dx,1) -- (0.2+3*\dx,1);
		\draw[line width=1.0] (0.3+3*\dx,0) -- (0.4+3*\dx,0) -- (0.4+3*\dx,1) -- (0.3+3*\dx,1);
		\node at (0.25+3*\dx,0.8) {\large $1$};
		\node at (0.6+3*\dx,0.52) {\Large $\mathbf{\dot{u}}$};
		
		\node[right] at (1.0, -0.5) {(b) Resistor as shunt};

		\def\dy{-2.2};
		\foreach \i in {0, 1, 2}
		{
			\draw[line width=1.0] (0.1+\i*\dx,0+\dy) -- (0+\i*\dx,0+\dy) -- (0+\i*\dx,1+\dy) -- (0.1+\i*\dx,1+\dy);
			\draw[line width=1.0] (1.0+\i*\dx,0+\dy) -- (1.1+\i*\dx,0+\dy) -- (1.1+\i*\dx,1+\dy) -- (1.0+\i*\dx,1+\dy);
		}
		\node at (1.5,0.5+\dy) {\Large $\mathbf{\ddot{z}}+$};
		\node at (1.5+\dx,0.5+\dy) {\Large $\mathbf{\dot{z}}+$};
		\node at (1.55+2*\dx,0.47+\dy) {\Large $\mathbf{z}=$};
		\draw[dash dot,line width=1.5, color=blue] (0.1,0.9+\dy) -- (1.,0.1+\dy);
		\draw[dashed,line width=1.5, color=orange] (0.1,0.7+\dy) -- (0.8,0.1+\dy);
		\draw[line width=1, color=blue, fill=gray!10] (0.1+\dx, 0.9+\dy) -- (1.0+\dx, 0.9+\dy) -- (1.0+\dx, 0.1+\dy) -- cycle;
		\draw[line width=1, color=blue, fill=gray!10] (0.1+2*\dx, 0.9+\dy) -- (1.0+2*\dx, 0.9+\dy) -- (1.+2*\dx, 0.1+\dy) -- cycle;
		\draw[line width=1.0] (0.2+3*\dx,0+\dy) -- (0.1+3*\dx,0+\dy) -- (0.1+3*\dx,1+\dy) -- (0.2+3*\dx,1+\dy);
		\draw[line width=1.0] (0.3+3*\dx,0+\dy) -- (0.4+3*\dx,0+\dy) -- (0.4+3*\dx,1+\dy) -- (0.3+3*\dx,1+\dy);
		\node at (0.25+3*\dx,0.8+\dy) {\large $1$};
		\node at (0.6+3*\dx,0.52+\dy) {\Large $\mathbf{\ddot{u}}$};
		
		\node[right] at (1.0, -0.5+\dy) {(c) Inductor as shunt};
		
	\end{tikzpicture}
	\caption{Illustration of matrix shapes of the second-order differential equations \protecting{\eqref{eq:rlc_second_order_1}}.}
	\label{fig:matrix_shapes}
\end{figure}

If we have equal component values, i.e. 
\begin{align*}
	r_{1} =&~ r_{2} = \ldots = r_{N} \text{,} \\
	c_{1} =&~ c_{2} = \ldots = c_{N} \quad \text{and} \\
	\ell_{1} =&~ \ell_{2} = \ldots = \ell_{N}
\end{align*}
then we can simplify our modeling approach. The triangular matrices for each ladder type, Eq. \eqref{eq:triangular_matrix}, are filled with ones and scaled as
\begin{equation*}
	\mathbf{T} = \left(a~b\right) ~ \mathbf{\tilde{T}} \quad \text{with} \quad 
	\mathbf{\tilde{T}} = \left. \mathbf{T} \right\rvert_{a_{n}=1,b_{n}=1} =
	\begin{pmatrix}
		1 & \ldots & 1 \\
		& \ddots & \vdots \\
		& & 1
	\end{pmatrix} 
	\textbf{.}
\end{equation*}
The multipliers of the scaling $a\cdot b$ are listed in Table \ref{table:triangular_matrix_values}. Hence, we yield the differential equations
\begin{subequations}
\begin{equation}
	\underbrace{(\ell~c)~ \mathbf{\tilde{T}}}_{=\mathbf{M}} ~ \ddot{z}(t) + \underbrace{(r~c)~ \mathbf{\tilde{T}}}_{=\mathbf{D}}~ \dot{z}(t) + \underbrace{\mathbf{F}}_{=\mathbf{S}}~ z(t) = G~\mu(t)
\end{equation}
for ladder type 1 (capacitor as shunt) with $\mu(t) = u(t)$, 
\begin{equation}
	\left(\frac{\ell}{r}\right)~ \mathbf{\tilde{T}}~ \ddot{z}(t) + \mathbf{F}~\dot{z}(t) + \left(\frac{1}{r~c}\right)~ \mathbf{\tilde{T}}~ z(t) = G~\mu(t)
\end{equation}
for ladder type 2 (resistor as shunt) with $\mu(t) = \partial_{t}u(t)$ and
\begin{equation}
	\mathbf{F}~ \ddot{z}(t) + \left(\frac{r}{\ell}\right) ~ \mathbf{\tilde{T}} ~ \dot{z}(t) + \left(\frac{1}{\ell~c}\right) ~ \mathbf{\tilde{T}} ~ z(t) = G~\mu(t)
\end{equation}
\label{eq:rlc_second_order_equal_coeff}
\end{subequations}
for ladder type 3 (inductor as shunt) with $\mu(t) = \partial_{t}^2 u(t)$. 



\subsection{First-Order System}
\label{sec:first_order_system}
Subsequently, we reformulate Eq. \eqref{eq:rlc_second_order_1} to yield a first-order system in the common state space formulation
\begin{align*}
	\dot{x}(t) =&~ \mathbf{A}~ x(t) + \mathbf{B}~\mu(t)\\
	y(t) =&~ \mathbf{C}^\top~x(t)
\end{align*}
with states $x:[0,T]\rightarrow \mathbb{R}^{2N}$, control input signal $\mu:[0,T]\rightarrow \mathbb{R}$, output signal $y:[0,T]\rightarrow \mathbb{R}$, system matrix $\mathbf{A} \in \mathbb{R}^{2N\times2N}$, and vectors $\mathbf{B}, \mathbf{C} \in \mathbb{R}^{2N}$. Hence, we firstly find
\begin{equation}
	\ddot{z}(t) + \mathbf{M}^{-1}~\mathbf{D} ~ \dot{z}(t) + \mathbf{M}^{-1}~\mathbf{S} ~ z(t) = \mathbf{M}^{-1}~\mathbf{G} ~ \mu(t)	 \label{eq:rlc_second_order_2}
\end{equation}
where we specify the matrices 
\begin{equation}
\mathbf{A_{I}} = -\mathbf{M}^{-1}~\mathbf{S} \quad \text{and} \quad \mathbf{A_{II}} := -\mathbf{M}^{-1}~\mathbf{D}	\label{eq:first_order_submatrices}
\end{equation}
and vector $\mathbf{\tilde{G}}:= \mathbf{M}^{-1}~\mathbf{G}$, and we note secondly 
\begin{equation*}
	\ddot{z}(t) = \mathbf{A_{I}} ~ z(t)  + \mathbf{A_{II}} ~ \dot{z}(t) + \mathbf{\tilde{G}} ~ z(t) \text{.}
\end{equation*}
We introduce the new states $x(t) = \left[x_{1}(t),x_{2}(t)\right]$ with $x_{1}(t) = z(t)$, $x_{2}(t) = \partial_{t} z(t)$, and we define the voltage at the last shunt as our output signal 
\begin{equation*}
	y(t) := z_{N}(t) = \left[\underbrace{0,\ldots,0}_{(N-1)\text{ zeros}},1,\overbrace{0\ldots,0}^{N \text{ zeros}}\right] 	\begin{pmatrix}
		{x}_{1} \\	
		{x}_{2}
	\end{pmatrix} = \mathbf{C}^{\top} x(t) \text{.} 
	\label{eq:state_space_output_signal}
\end{equation*}

\begin{marginfigure}
	\centering
	\begin{tikzpicture}[scale=1,line cap=round,line join=round,x=1.0cm,y=1.0cm]	
		\def\dx{2};
		\def\dy{-2.4};
		
		\foreach \i in {1, 2, 3}
		{
			\draw[line width=1.0] (0.1+\i*\dx,0+\dy) -- (0+\i*\dx,0+\dy) -- (0+\i*\dx,1+\dy) -- (0.1+\i*\dx,1+\dy);
			\draw[line width=1.0] (1.0+\i*\dx,0+\dy) -- (1.1+\i*\dx,0+\dy) -- (1.1+\i*\dx,1+\dy) -- (1.0+\i*\dx,1+\dy);
		}

		\draw[line width=1, color=blue, fill=gray!10] (0.1+\dx, 0.9+\dy) -- (1.0+\dx, 0.9+\dy) -- (1.0+\dx, 0.1+\dy) -- cycle;
		
		\draw[dash dot,line width=1.5, color=blue] (0.1+2*\dx,0.9+\dy) -- (1.+2*\dx,0.1+\dy);
		\draw[dashed,line width=1.5, color=orange] (0.1+2*\dx,0.7+\dy) -- (0.8+2*\dx,0.1+\dy);
		
		\draw[dash dot,line width=1.5, color=blue] (0.1+3*\dx,0.9+\dy) -- (1.+3*\dx,0.1+\dy);
		\draw[dashed,line width=1.5, color=orange] (0.1+3*\dx,0.7+\dy) -- (0.8+3*\dx,0.1+\dy);
		\draw[dashed,line width=1.5, color=orange] (0.3+3*\dx,0.9+\dy) -- (1.0+3*\dx,0.3+\dy);
		
		\node[right] at (1.1+\dx, 1.1+\dy) {\large \textbf{-1}};		
		\node[right] at (1.4+\dx, 0.5+\dy) {\Huge \textbf{.} };		
		\node[right] at (1.3+2*\dx, 0.4+\dy) {\Large = };
		
		\node[right] at (0.2+\dx, -0.6+\dy) {(a) Tridiagonal Matrix $\mathbf{L}$};		
		
		\foreach \i in {1, 2, 3}
		{
			\draw[line width=1.0] (0.1+\i*\dx,0+2*\dy) -- (0+\i*\dx,0+2*\dy) -- (0+\i*\dx,1+2*\dy) -- (0.1+\i*\dx,1+2*\dy);
			\draw[line width=1.0] (1.0+\i*\dx,0+2*\dy) -- (1.1+\i*\dx,0+2*\dy) -- (1.1+\i*\dx,1+2*\dy) -- (1.0+\i*\dx,1+2*\dy);
		}
		\draw[line width=1, color=blue, fill=gray!10] (0.1+\dx, 0.9+2*\dy) -- (1.0+\dx, 0.9+2*\dy) -- (1.0+\dx, 0.1+2*\dy) -- cycle;
		
		\draw[line width=1, color=blue, fill=gray!10] (0.1+2*\dx, 0.9+2*\dy) -- (1.0+2*\dx, 0.9+2*\dy) -- (1.0+2*\dx, 0.1+2*\dy) -- cycle;
		
		\fill[fill=gray!10] (0.1+3*\dx, 0.9+2*\dy) -- (1.0+3*\dx, 0.9+2*\dy) -- (1.0+3*\dx, 0.1+2*\dy) -- cycle;
		\draw[dashed, line width=1, color=blue] (0.1+3*\dx, 0.9+2*\dy) -- (1.0+3*\dx, 0.9+2*\dy) -- (1.0+3*\dx, 0.1+2*\dy);
		\draw[solid, line width=1, color=blue] (0.1+3*\dx, 0.9+2*\dy) -- (1.0+3*\dx, 0.1+2*\dy);
		
		\node[right] at (1.1+\dx, 1.1+2*\dy) {\large \textbf{-1}};		
		\node[right] at (1.4+\dx, 0.5+2*\dy) {\Huge \textbf{.} };		
		\node[right] at (1.3+2*\dx, 0.4+2*\dy) {\Large = };
		
		\node[right] at (0.2+\dx, -0.6+2*\dy) {(b) Diagonal or Triangular Matrix $\mathbf{P}$};

		\foreach \i in {1, 2, 3}
		{
			\draw[line width=1.0] (0.1+\i*\dx,0+3*\dy) -- (0+\i*\dx,0+3*\dy) -- (0+\i*\dx,1+3*\dy) -- (0.1+\i*\dx,1+3*\dy);
			\draw[line width=1.0] (1.0+\i*\dx,0+3*\dy) -- (1.1+\i*\dx,0+3*\dy) -- (1.1+\i*\dx,1+3*\dy) -- (1.0+\i*\dx,1+3*\dy);
		}

		\draw[dash dot,line width=1.5, color=blue] (0.1+\dx,0.9+3*\dy) -- (1.0+\dx,0.1+3*\dy);
		\draw[dashed,line width=1.5, color=orange] (0.1+\dx,0.7+3*\dy) -- (0.8+\dx,0.1+3*\dy);
		
		\draw[line width=1, color=blue, fill=gray!10] (0.1+2*\dx, 0.9+3*\dy) -- (1.0+2*\dx, 0.9+3*\dy) -- (1.0+2*\dx, 0.1+3*\dy) -- cycle;

		\draw[line width=1.2, color=blue] (0.1+3*\dx, 0.9+3*\dy) -- (0.1+3*\dx, 0.1+3*\dy);
		\foreach \i in {1, ..., 8}
		{
			\draw[line width=1.2, color=blue] (0.1+\i*0.11+3*\dx, 0.9-\i*0.1+3*\dy) -- (0.1+\i*0.11+3*\dx, 0.1+3*\dy);
			\foreach \j in {1, ..., \i}
			{
				\fill [line width=1.0, color=orange] (0.1+\i*0.11 +3*\dx, 1.0 -0.1*\j +3*\dy) circle [radius = 0.025];	
			}
		}

		\node[right] at (1.1+\dx, 1.1+3*\dy) {\large \textbf{-1}};		
		\node[right] at (1.4+\dx, 0.5+3*\dy) {\Huge \textbf{.} };		
		\node[right] at (1.3+2*\dx, 0.4+3*\dy) {\Large = };
		
		\node[right] at (0.2+\dx, -0.6+3*\dy) {(c) Dense Matrix $\mathbf{Q}$};

	\end{tikzpicture}
	\caption{Matrix multiplications to yield sub-matrices $\mathbf{A_{I}}$ and $\mathbf{A_{II}}$ in the state-space model.}
	\label{fig:matrix_multiplication}
\end{marginfigure}
So, we formulate the state space as 
\begin{subequations}
\begin{align}
	\begin{pmatrix}
		\dot{x}_{1} \\	
		\dot{x}_{2}
	\end{pmatrix}
	=&~
	\underbrace{
		\begin{pmatrix}
			\mathbf{0} & \mathbf{I} \\
			\mathbf{A_{I}} & \mathbf{A_{II}}
		\end{pmatrix}
	}_{=: \mathbf{A}}
	\begin{pmatrix}
		{x}_{1} \\	
		{x}_{2}
	\end{pmatrix}
	+
	\underbrace{
		\begin{pmatrix}
			0  \\
			\mathbf{\tilde{G}}
		\end{pmatrix}
	}_{=: \mathbf{B}}
	\mu(t) \label{eq:state_space1}
	\text{,} \\
	y(t) =&~ \mathbf{C}^\top ~ x(t) \text{.} \label{eq:state_space2}
\end{align}
\label{eq:state_space}
\end{subequations}
In Eq. \eqref{eq:state_space1}, in matrix $\mathbf{A}$ we assume a zero sub-matrix $\mathbf{0} = \mathbf{0}_{N\times N}$ and in vector $\mathbf{B}$ we mean a row vector $\mathbf{0} = \mathbf{0}_{N}$.
If matrix $\mathbf{M}$ in Eq. (\ref{eq:rlc_second_order_1},\ref{eq:rlc_second_order_2}) is not well conditioned, i.e. $M$ has a large condition number $\kappa = \frac{\max_{i} |\lambda_{i}|}{\min_{i} |\lambda_{i}|} \gg 1$ with eigenvalues $\lambda_{i}$ and $i \in \{1,\ldots,2N\}$, then a numerical computation of sub-matrices $A_{\mathrm{I}}$ and $A_{\mathrm{II}}$ as in Eq. \eqref{eq:first_order_submatrices} might be prone to numerical errors. Hence, we find a inverse matrix $\mathbf{M}^{-1}$ and the concluding multiplications, Eq. \eqref{eq:first_order_submatrices}, in a closed form. We have matrix $\mathbf{M}$ either in a triangular shape, see Eq. \eqref{eq:triangular_matrix}, or as a finite difference stencil, see Eq. \eqref{eq:finite_diff_stencil}. Thus, we have three types of matrix multiplications as sketched in  Fig. \ref{fig:matrix_multiplication} and we derive them subsequently.

\subsection*{Inverse Triangular Matrix}
To compute sub-matrices $\mathbf{A_{I}}$ and $\mathbf{A_{II}}$ for circuit types (a) and (b), we have matrix $\mathbf{M}$ in a triangular form $\mathbf{T}$, see Eq. \eqref{eq:triangular_matrix}. We find the inverse of $\mathbf{T}$ as the upper finite difference stencil shape 
\begin{align*}
	\mathbf{T}^{-1} = 
	\begin{pmatrix}
		\frac{1}{a_{1} ~ b_{1}} & \frac{-1}{a_{2} ~ b_{1}} & 0  						& \cdots & 0 \\[1ex]
		0 						 &   \frac{1}{a_{2} ~ b_{2}} & \frac{-1}{a_{3} ~ b_{2}} &  & \vdots \\
		\vdots 				 		 &  \ddots 					 & \ddots 					 & \ddots & 0	 \\
		&  & 0 & \frac{1}{a_{N-1} ~ b_{N-1}} & \frac{-1}{a_{N} ~ b_{N-1}} \\[1ex]
		0  & \cdots 	& &  			0			   & \frac{1}{a_{N} ~ b_{N}} 
	\end{pmatrix} \text{.}
\end{align*}

Here, we distinguish two scenarios: a multiplication of $\mathbf{T}^{-1}$ with a finite difference stencil $\mathbf{F}$, see Eq. \eqref{eq:finite_diff_stencil}, and with triangular matrix $\mathbf{T}$. In the first case, we define the resulting matrix as
\begin{equation}
	\mathbf{L} := -\mathbf{T}^{-1} ~\cdot~ \mathbf{F}
\end{equation}
and we see that it is a tridiagonal matrix\sidenote{Matrix $\mathbf{L}$ has the shape of a second-order difference stencil, which approximates the one-dim. Laplace operator.} as
\begin{equation}
	\mathbf{L} =
	\begin{pmatrix}
		\gamma_{1} & \beta_{1} \\
		\alpha_{2} & \gamma_{2} & \beta_{2} \\
		& \ddots & \ddots & \ddots \\
		& & \alpha_{N-1} & \gamma_{N-1} & \beta_{N-1} \\
		& & & \alpha_{N} & -\alpha_{N}
	\end{pmatrix}
	\label{eq:laplace_matrix}
\end{equation}
with sub-diagonal entries
\begin{equation*}
	\alpha_{n} = \frac{1}{a_{n}~ b_{n}} ~ \text{,} \quad 
	\beta_{n} =  \frac{1}{a_{n+1}~ b_{n}} \\
\end{equation*}
and diagonal elements
\begin{equation*}
	\gamma_{n} = -\left[\alpha_{n}+\beta_{n}\right] =  \frac{-1}{a_{n}~ b_{n}} +  \frac{-1}{a_{n+1}~ b_{n}}
\end{equation*}
for $n\in\{1,\ldots,N\}$. Matrix shape $\mathbf{L}$ is portrayed in Fig. \ref{fig:matrix_multiplication} (a).

In the second case, we have the second upper triangular matrix as $\hat{\mathbf{T}} := \left. \mathbf{T} \right\rvert_{a_n=\tilde{a}_n,b_n=\tilde{b}_n}$ with coefficients $\tilde{a}_{n}$ and $\tilde{b}_{n}$ for $n\in\{1,\ldots,N\}$. We define the product as 
\begin{equation*}
	\mathbf{P} := - \mathbf{T}^{-1} ~ \cdot ~ \hat{\mathbf{T}}
\end{equation*}
which has a triangular form again, see Fig. \ref{fig:matrix_multiplication} (b), as
\begin{equation}
	\mathbf{P} = 
	\begin{pmatrix}
		\alpha_{1} & \beta_{1,2} &  \beta_{1,3} & \ldots &   \beta_{1,N} \\
		& \alpha_{2} & \beta_{2,3} &  \ldots &  \beta_{2,N} \\
		& & \ddots & \vdots \\
		& & & \alpha_{N-1} & \beta_{N-1,N} \\
		& & & & \alpha_{N}
	\end{pmatrix}
	\label{eq:triangular_P}
\end{equation}
with entries
\begin{equation}
	\alpha_{n} = -\frac{\tilde{a}_{n}}{a_{n}} \quad \text{and} \quad \beta_{n,k} = -\left[\frac{\tilde{a}_{n}}{a_{n}} - \frac{\tilde{a}_{n+1}}{a_{n+1}} \right] \frac{\tilde{b}_{k}}{b_{n}} \text{.}
\end{equation} 
In Table \ref{table:triangular_matrix_values}, we see that the coefficients $b_{n}$ are equal for both triangular matrices and so we have $\tilde{b}_{k} \equiv b_{k}$ for $k \in \{1,\ldots,N\}$. The coefficients $a_{n}$ and $\tilde{a}_{n}$ differ in general, but if they are scaled\sidenote{We have such a case if all values of the related electrical elements are equals as described in the end of this section.} as  $\tilde{a}_{n} = \epsilon ~ a_{n}$ for all $n \in \{1,\ldots,N\}$ with $\epsilon \in \mathbb{R}$, then we yield a pure diagonal matrix
\begin{equation}
	\mathbf{P} =~ -\epsilon~\mathbf{I} =~  -\frac{\tilde{a}}{a}
	\begin{pmatrix}
		1 \\
		& 1 \\
		& & \ddots \\
		& & & 1
	\end{pmatrix} \text{.} \label{eq:triangular_P_id_matrix}
\end{equation}
with identity matrix $\mathbf{I} =\operatorname{diag}(1,\ldots,1)$. The input signal $\mu$ is scaled by vector $\mathbf{B} = [\mathbf{0}, \mathbf{\tilde{G}}]^\top$ and we find $\mathbf{\tilde{G}}$ as 
\begin{equation*}
	\mathbf{\tilde{G}} ~=~ \mathbf{T}^{-1} ~ \mathbf{G} = \mathbf{T}^{-1} ~ 
	\begin{pmatrix}
		1 \\
		0 \\
		\vdots \\
		0
	\end{pmatrix} ~=~
	\begin{pmatrix}
		\frac{1}{a_{1} ~ b_{1}} \\
		0 \\
		\vdots \\
		0
	\end{pmatrix}
\end{equation*}
and further, we yield
\begin{equation*}
	\mathbf{B} = 
	\begin{pmatrix}
		0 \\
		\vdots \\
		0 \\
		\frac{1}{a_{1} ~ b_{1}} \\
		0 \\
		\vdots \\
		0
	\end{pmatrix}
	\begin{matrix}
		\\
		\\
		\\
		\leftarrow (N+1)\text{-th position}\\
		\\
		\\
		~
	\end{matrix}
	\textbf{.}
\end{equation*}

\subsection*{Inverse Finite Difference Stencil}

In case of ladder type 3, we need to evaluate another matrix multiplication. Here, we invert the finite difference stencil $\tilde{F}$ and we obtain a lower triangular matrix
\begin{equation*}
	\mathbf{F}^{-1} =
	\begin{pmatrix}
		1 & 0  &  0\\
		\vdots 	& \ddots &  0 \\
		1 &	\cdots	 &     1 
	\end{pmatrix} \text{.}
\end{equation*}
We multiply this lower triangular matrix with $\mathbf{T}$ as in Eq. \eqref{eq:triangular_matrix} and we define the result as 
\begin{equation*}
	\mathbf{Q} := - \mathbf{F}^{-1} ~\cdot~ \mathbf{T}
\end{equation*}
where $\mathbf{Q} = [q_{n,k}]$ with entries
\begin{equation*}
	q_{n,k} = - b_{k} \sum_{i=1}^{\min\{n,k\}} a_{i}
\end{equation*}
for $n,k \in \{1,\ldots,N\}$. The resulting matrix shape is illustrated in Fig. \ref{fig:matrix_multiplication} (c). If all values of $b_{k}$ are equal as
\begin{equation*}
	b_{1}=\ldots=b_{N}~=:~\tilde{b}	\text{,}
\end{equation*}
then this matrix multiplication results in a symmetric shape
\begin{equation}
	\mathbf{Q} =~ 
	\def\aa{\color{juliapurple}\tilde{q}_{2}}
	\def\ab{\color{juliablue}\tilde{q}_{N-1}}
	\begin{pmatrix}
		\tilde{q}_{1} & \tilde{q}_{1} & \tilde{q}_{1} & \ldots & \tilde{q}_{1} & \tilde{q}_{1} \\
		\tilde{q}_{1} & \aa & \aa & \ldots & \aa & \aa \\
		& \aa & \tilde{q}_{3} & \ldots & \tilde{q}_{3} & \tilde{q}_{3} \\
		\vdots     & \vdots & \vdots & \ddots \\
		\tilde{q}_{1}& \aa & \tilde{q}_{3} & & \ab & \ab \\
		\tilde{q}_{1} & \aa & \tilde{q}_{3} & & \ab & \tilde{q}_{N}
	\end{pmatrix}
	\label{eq:dense_matrix}
\end{equation}
with $\tilde{q}_{n} = -\tilde{b} ~ \sum\limits_{i=1}^{n} a_{i}$, see Fig. \ref{fig:dense_matrix} for illustration.
\begin{marginfigure}
	\centering
	\begin{tikzpicture}[scale=1,line cap=round,line join=round,x=1.0cm,y=1.0cm]	
		\node at (-1,1) {\huge $\mathbf{Q}=$ };
		\draw[line width=1] (0.3, -0.05) -- (0, -0.05) -- (0, 2.05) -- (0.3, 2.05);
		\draw[line width=1] (1.9, -0.05) -- (2.2, -0.05) -- (2.2, 2.05) -- (1.9, 2.05);
		
		\draw[line width=2,color=wong1] (0.2,0.1) -- (0.2,1.9) -- (2.0,1.9);

		\def\dx{0.2}
		\draw[line width=2,color=wong2] (0.2+\dx,0.1) -- (0.2+\dx,1.9-\dx) -- (2.0,1.9 -\dx);
		\draw[line width=2,color=wong3] (0.2+2*\dx,0.1) -- (0.2+2*\dx,1.9-2*\dx) -- (2.0,1.9 -2*\dx);
		\draw[line width=2,color=wong4] (0.2+3*\dx,0.1) -- (0.2+3*\dx,1.9-3*\dx) -- (2.0,1.9 -3*\dx);

		\draw[line width=2,color=wong5] (0.2+7*\dx,0.1) -- (0.2+7*\dx,1.9-7*\dx) -- (2.0,1.9 -7*\dx);
		\draw[line width=2,color=wong6] (0.2+8*\dx,0.1) -- (0.2+8*\dx,1.9-8*\dx) -- (2.0,1.9 -8*\dx);
		\fill [line width=1.0, color=juliapurple] (0.2+8.8*\dx,1.9-8.8*\dx) circle [radius = 0.06];	
		
		\fill [line width=1.0, color=blue] (0.2+4.3*\dx,1.9-4.3*\dx) circle [radius = 0.05];	
		\fill [line width=1.0, color=blue] (0.2+5*\dx,1.9-5*\dx) circle [radius = 0.05];	
		\fill [line width=1.0, color=blue] (0.2+5.7*\dx,1.9-5.7*\dx) circle [radius = 0.05];	
	\end{tikzpicture}
	\caption{Illustraion of symmetric dense matrix $\mathbf{Q}$ where common rows and columns have equal entries.}
	\label{fig:dense_matrix}
\end{marginfigure}

We note the scaling vector $B$ of input signal $\mu$ in this case as
\begin{equation*}
	\mathbf{B} = 
	\begin{pmatrix}
		0 \\
		\vdots \\
		0 \\
		1 \\
		1 \\
		\vdots \\
		1
	\end{pmatrix}
	\begin{matrix}
		\\
		\\
		\\
		\leftarrow (N+1)\text{-th position}\\
		\\
		\\
		~
	\end{matrix}
\end{equation*}
because we have here $\mathbf{\tilde{G}} = \mathbf{F}^{-1}~\mathbf{G} = \left[1,\ldots,1\right]^\top$.

\subsection*{Equal Component Parameters}

We consider equal component values as in the end of Section \ref{sec:state_space_model_second_order}, and we formulate the first-order differential equations analog to the second-order formulae, see Eq. \eqref{eq:rlc_second_order_equal_coeff}. Hence, we yield for Eq. \eqref{eq:laplace_matrix} and \eqref{eq:dense_matrix} the matrices $\mathbf{L} = (a~b)~\mathbf{\tilde{L}}$ and $\mathbf{Q} = (a~b)~\mathbf{\tilde{Q}}$ with 
\begin{fullwidth}
\begin{equation*}
	\mathbf{\tilde{L}} = 
	\begin{pmatrix}
		-2 & 1 \\
		1 & -2 & 1 \\
		& \ddots & 	\ddots & \ddots & \\
		& & 1 & -2 & 1 \\
		& & & 1 & -1
	\end{pmatrix} 
	\qquad \text{and} \qquad
	\mathbf{\tilde{Q}} =
	\begin{pmatrix}
		-1 & -1 & \ldots & -1 & -1 \\
		-1 & -2 & \ldots & -2 & -2 \\
		\vdots & \vdots & \ddots \\
		\vdots & \vdots & & -[N-1] & -[N-1] \\ 
		-1 & -2 & & -[N-1] & -N
	\end{pmatrix}
\end{equation*}	
\end{fullwidth}
and constant coefficient values $a$ and $b$. Furthermore, matrix $\mathbf{P}$ becomes a scaled identity matrix, see Eq. \eqref{eq:triangular_P} and \eqref{eq:triangular_P_id_matrix}. In the next step, we formulate the state space as in Eq. \eqref{eq:state_space1}. Here, we find the first row for all three circuit types as $\dot{x}_{1}(t) = x_{2}(t)$ and the second row as
\begin{subequations}
\begin{equation}
	\dot{x}_{2}(t) = \underbrace{\left(\frac{1}{\ell~c}\right) ~\mathbf{\tilde{L}}}_{=\mathbf{A_{I}}} ~ x_{1}(t) + 
	\underbrace{\left(-\frac{r}{\ell}\right) ~ \mathbf{I}}_{=\mathbf{A_{II}}} ~ x_{2}(t) + 
	\frac{1}{\ell~c} ~
	\begin{pmatrix}
		1\\
		0\\
		\vdots \\
		0
	\end{pmatrix}
	\mu(t)
\end{equation}		
for ladder type 1 (capacitor as shunt), 
\begin{equation}
	\dot{x}_{2}(t) = \left(\frac{-1}{\ell~c}\right) ~
	\mathbf{I} ~ 	x_{1}(t) + 
	\left(\frac{r}{\ell}\right) ~ \tilde{\mathbf{L}} ~ x_{2}(t) + 
	\left(\frac{r}{\ell}\right) ~
	\begin{pmatrix}
		1\\
		0\\
		\vdots \\
		0
	\end{pmatrix}
	\mu(t)
\end{equation}	
for ladder type 2 (resistor as shunt) and
\begin{equation}
	\dot{x}_{2}(t) = \left(\frac{1}{\ell~c}\right) ~ \mathbf{\tilde{Q}} ~ x_{1}(t) + 
	\left(\frac{r}{\ell}\right) ~ \mathbf{\tilde{Q}} ~ x_{2}(t) + 
	\begin{pmatrix}
		1\\
		1\\
		\vdots \\
		1
	\end{pmatrix}
	\mu(t)
\end{equation}	
for ladder type 3 (inductor as shunt). 
\label{eq:rlc_first_order_equal_coeff}
\end{subequations}
We remind that the system input $\mu$ in Eq. \eqref{eq:rlc_first_order_equal_coeff} relates for each ladder type to the circuit input $u$ as described in Eq. \eqref{eq:input_signal}.

\section{Simulation Examples}
\label{sec:simulation_examples}
\begin{margintable}
	\renewcommand{\arraystretch}{1.3}
	\centering
	\caption{\MakeUppercase{Simulation Parameters}}
	\begin{tabular}{l l l}
	\toprule
	& Val. & Description \\
	\midrule
	N & 10 & Num. of Circuits\\
	$r$ & 2 & Resistance in $[\Omega]$ \\
	$c$ & 3 & Capacitance in $[F]$\\
	$\ell$ & 5 & Inductance in $[H]$\\
	\bottomrule	
	\end{tabular}
	\label{table:simulation_parameters}
\end{margintable}
For each proposed ladder type, we assume an example consisting of ten sub-circuits. The resistor, capacitor and inductor values  are equal for all sub-circuits, they are listed in Table \ref{table:simulation_parameters}. So, we have the scaling factors $\frac{1}{\ell~c} = \frac{1}{15}$ and $\frac{r}{\ell} = \frac{2}{5}$ for Eq. \eqref{eq:rlc_first_order_equal_coeff}. We demonstrate the dynamical behavior of our ladder systems with two types of input signals. Firstly, we consider a step input and secondly a sweep signal. We remind, that the voltage at the last shunt is our output signal $y(t) = z_{10}(t)$, see Eq. \eqref{eq:state_space_output_signal}.


\subsection{Step Input Signal}

\begin{figure}[t!]
	\centering
	\subfloat[Step Signal $\mu$]{\includegraphics[width=0.95\columnwidth]{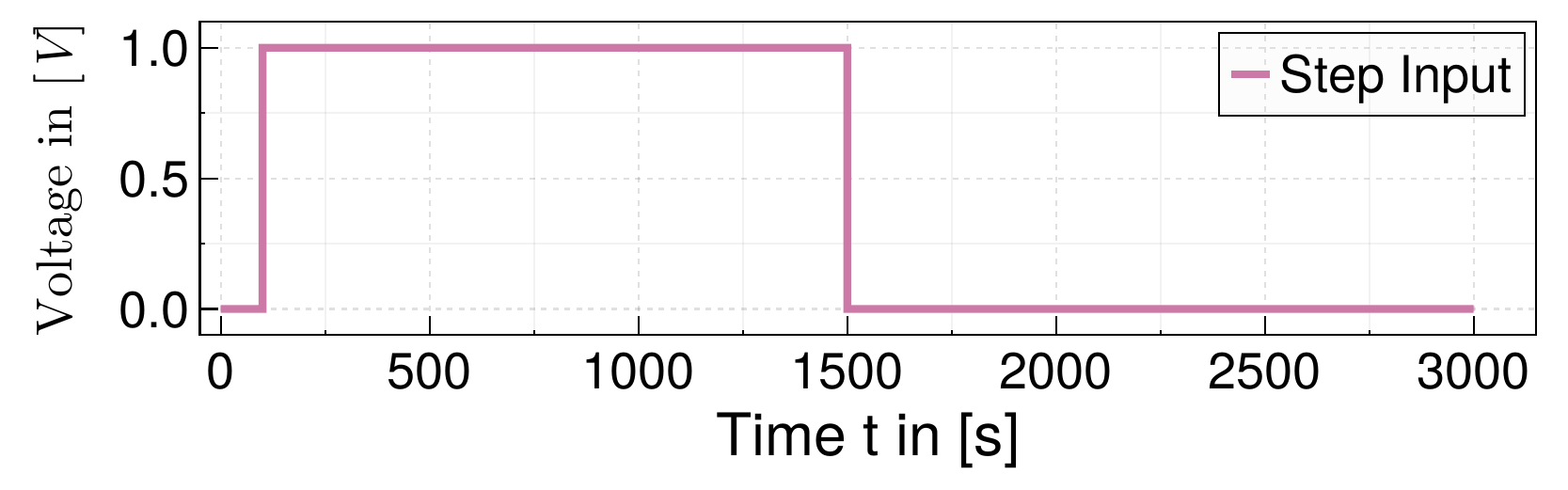}}
	
	\subfloat[Type 1: Capacitor as Shunt]{\includegraphics[width=0.95\columnwidth]{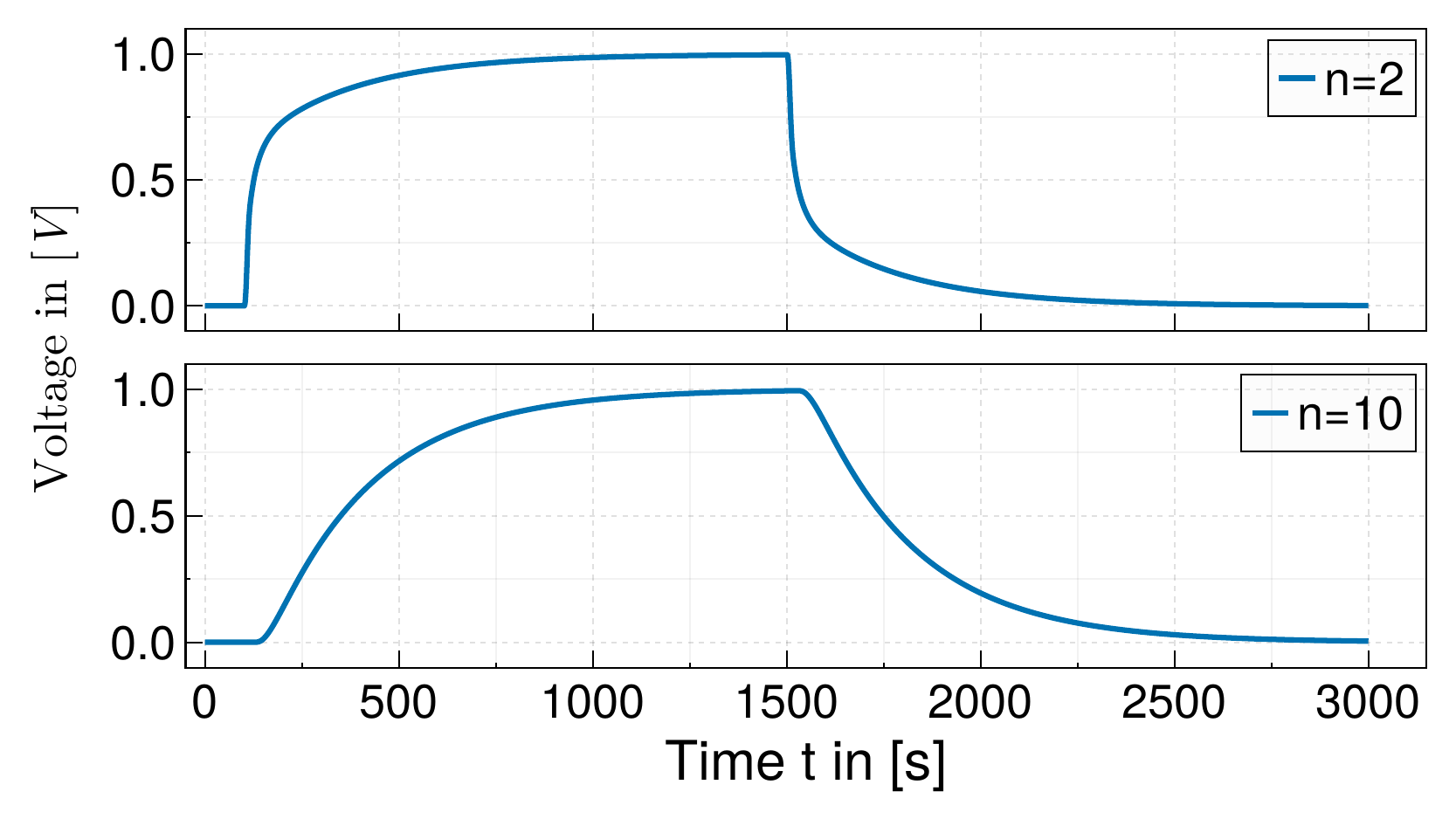}}
	
	\subfloat[Type 2: Resistor as Shunt]{\includegraphics[width=0.95\columnwidth]{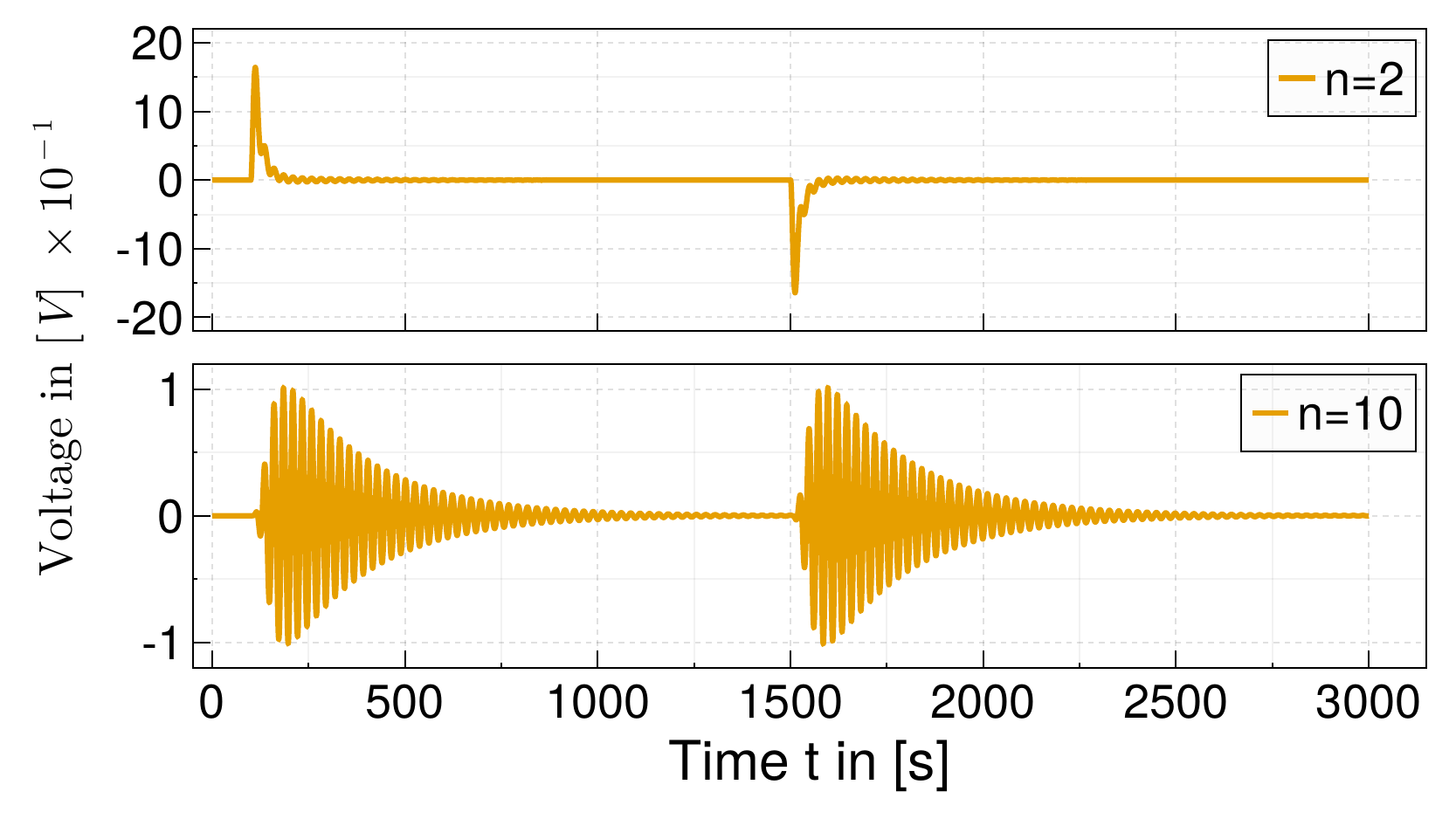}}
	
	\subfloat[Type 3: Inductor as Shunt]{\includegraphics[width=0.95\columnwidth]{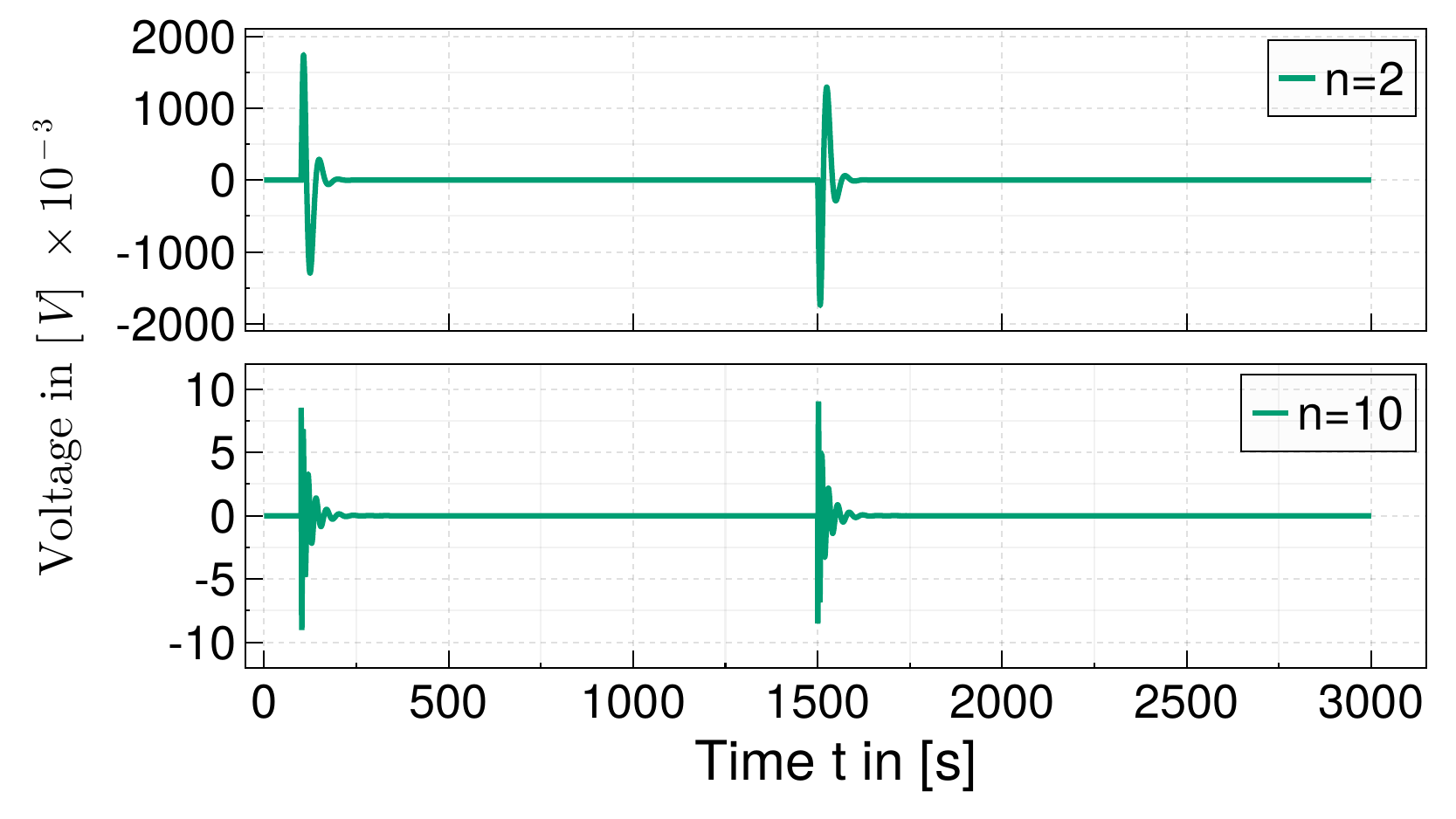}}
	
	\caption{Step input signal in (a) and the resulting voltages at position $n\in\{2,10\}$ of all three ladder types in (b) to (d).}
	\label{fig:step_input_output_results}
\end{figure}

In the first example, we assume a step input signal
\begin{equation}
	\mu(t) = 
	\begin{cases}
		1 & ~\text{for } t \in [50, 1500] \text{,} \\
		0 & ~ \text{for } t \in [0,50) \cup (1500,3000]
	\end{cases}
\end{equation}
with time t in $[s]$ and we compute the response for three state space models \eqref{eq:rlc_first_order_equal_coeff}.\sidenote{We remark that the original circuit input $u(t)$ is the first and second anti-derivative of this step signal in case of ladder type 2 and 3, resp.} The input signal and the simulation results are visualized in Fig. \ref{fig:step_input_output_results}. We find a charging and discharging dynamics in case of ladder type 1 in Fig. \ref{fig:step_input_output_results} (b), and oscillations for ladder type 2 and 3 in Fig. \ref{fig:step_input_output_results} (c), (d). Comparing the oscillations, we have a long and intensive oscillation for ladder type 2 in the last shunt, while the step response drives quickly to zero for type 3. 

\subsection{Symmetric Sweep Input Signal}
In this second example, we assume a symmetric sweep signal as circuit input $u(t)$, which starts and terminates at zero Volt. Hence, in case of ladder type 2 and 3, see Fig. \ref{fig:cascaded_rlc_circuits} (b) and (c), we need to differentiate the input voltage once and twice, see Eq. \eqref{eq:input_signal}. The two-sided chirp signal is constructed in two steps: we assume an oscillation
\begin{figure*}[t!]
	\centering
	\subfloat[Input Signals]{\includegraphics[width=0.47\columnwidth]{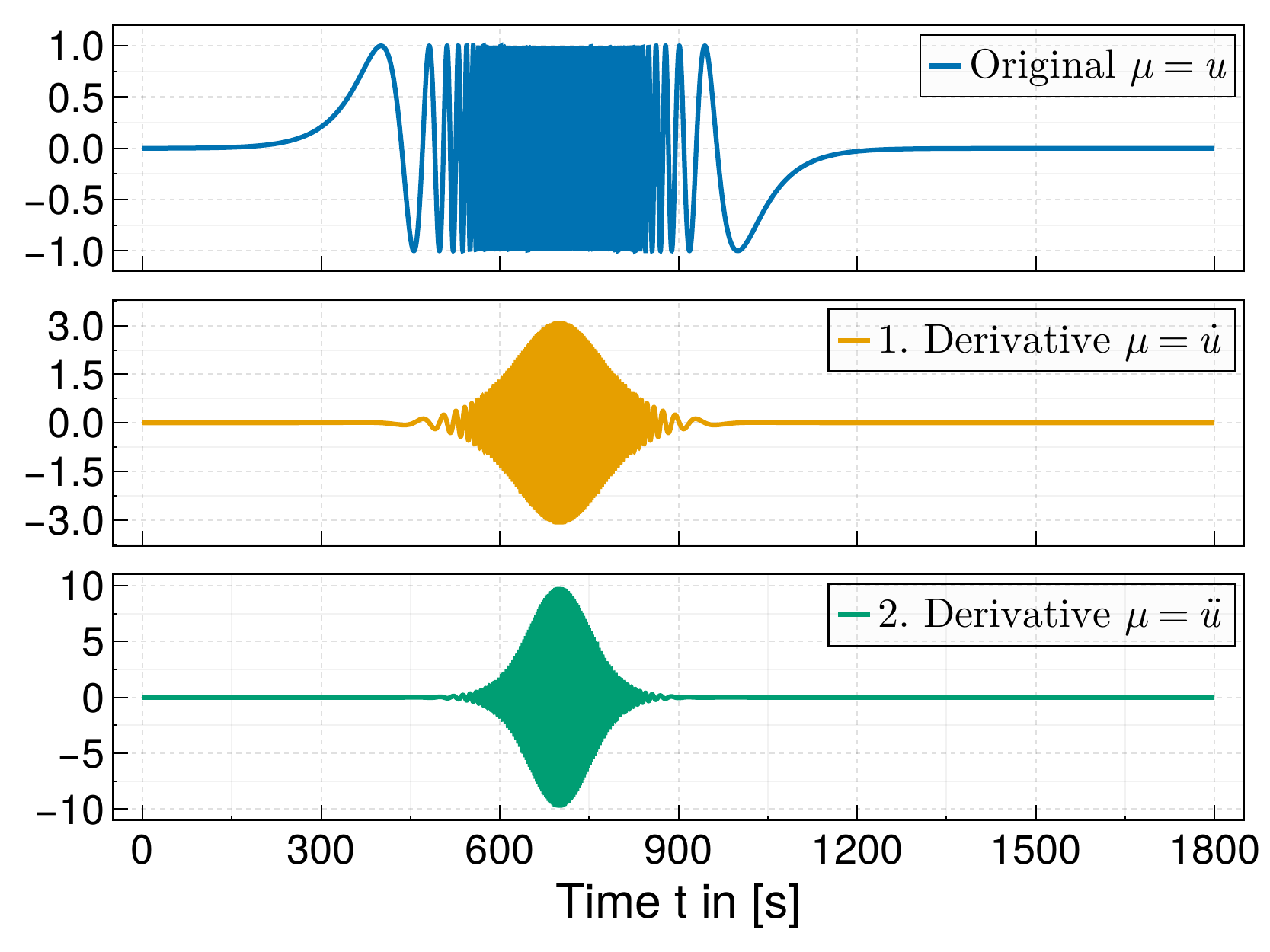}}
	\hfill
	\subfloat[Type 1: Capacitor as Shunt]{\includegraphics[width=0.47\columnwidth]{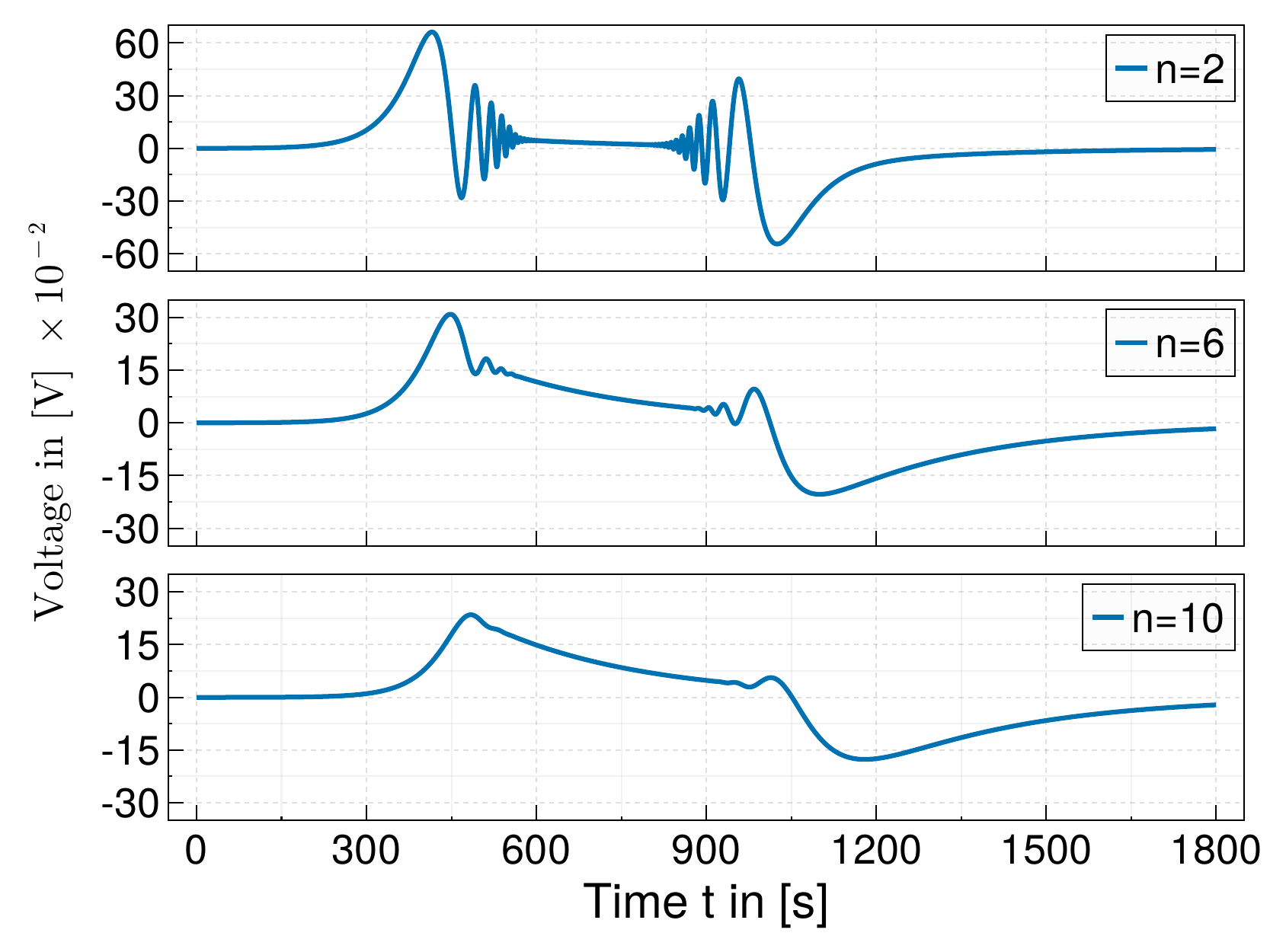}}
	
	\subfloat[Type 2: Resistor as Shunt]{\includegraphics[width=0.47\columnwidth]{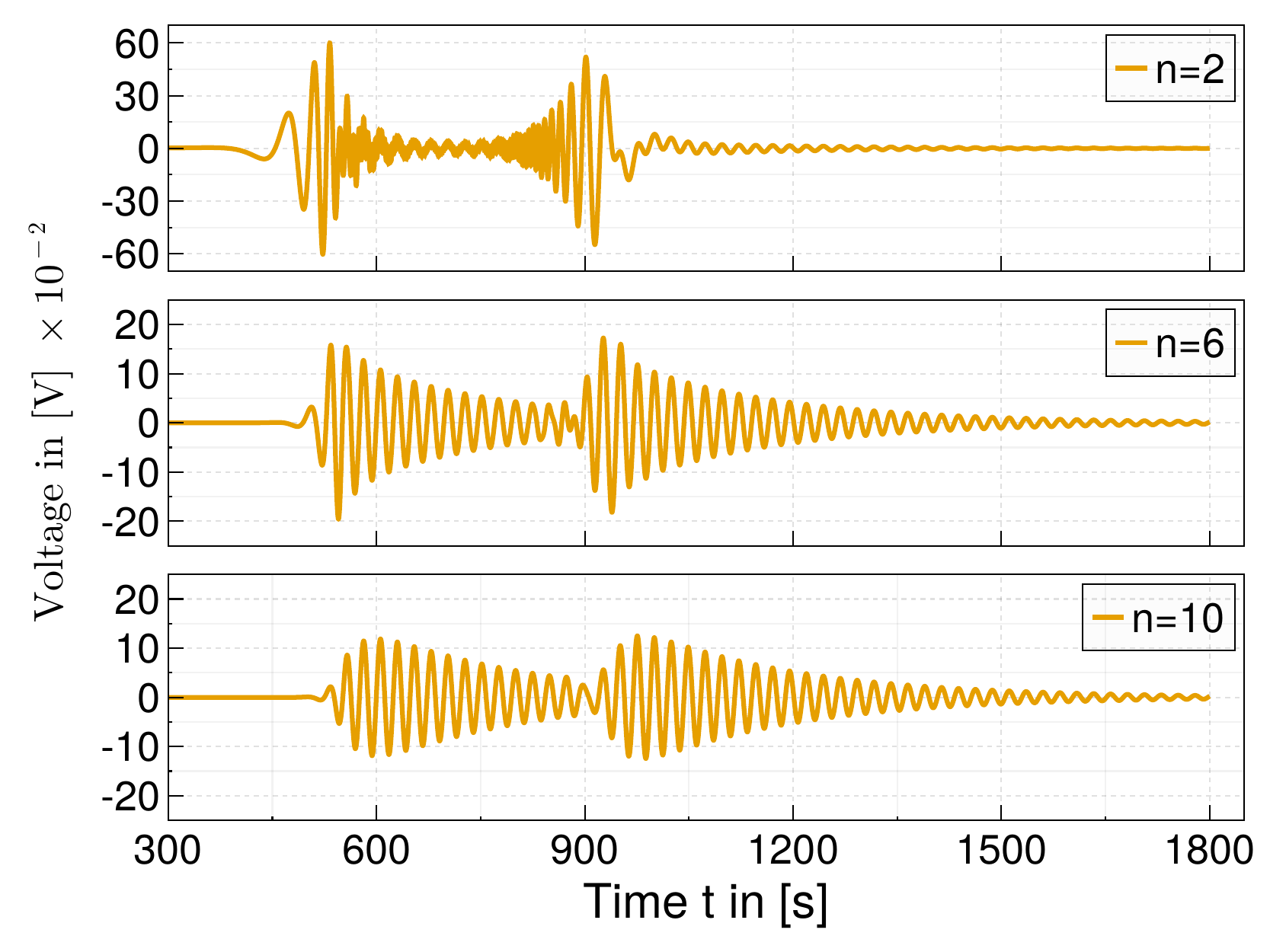}}
	\hfill
	\subfloat[Type 3: Inductor as Shunt]{\includegraphics[width=0.47\columnwidth]{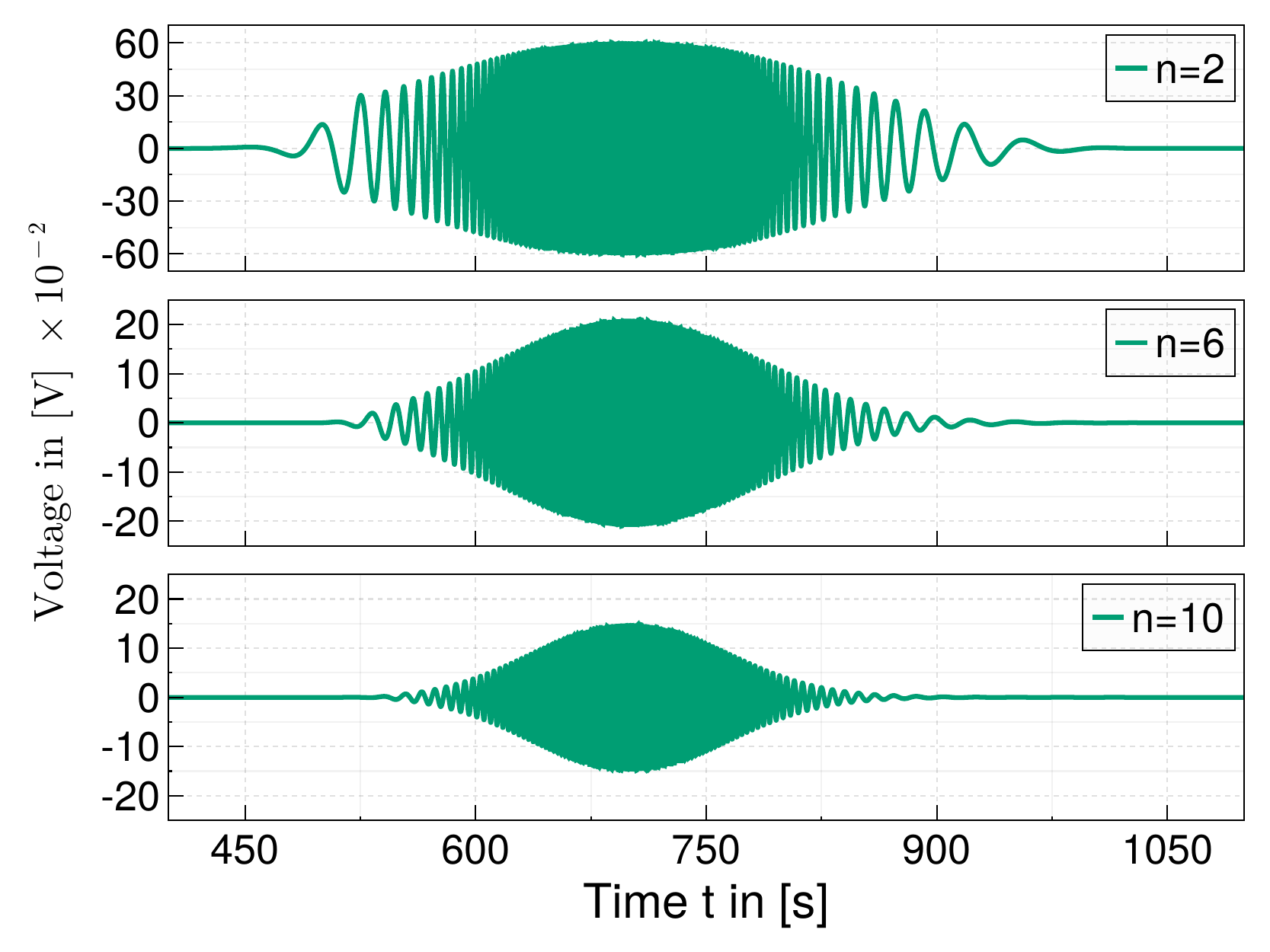}}
	
	\hfill
	
	\hfill
	
	\caption{Symmetric sweep input signal and its derivatives in (a) and the resulting voltages at position $n\in\{2,6,10\}$ of all three ladder types in (b) to (d).}
	\label{fig:sweep_results}
\end{figure*}	
\begin{equation}
	f(t,p) = \sin\left(p~\pi~g(t,p)\right) \label{eq:input_base_oscillation}
\end{equation}
with an embedded function
\begin{equation*}
	g(t,p) := \tanh\left(\frac{t}{p}\right) \text{,}
\end{equation*}
and we choose parameter $p=100$ and time shift $t \rightarrow [t-700]$ to yield the original voltage signal
\begin{equation*}
	u(t) = f(t-700,100) \textbf{.}
\end{equation*}
Next, we formulate the first and second derivative of $u(t)$. So, we differentiate symbolically the oscillation function, Eq. \eqref{eq:input_base_oscillation}. The first derivative results in 
\begin{equation*}
	\partial_{t} f(t,p) = ~ \pi~\cos\left(p~\pi~g(t,p)\right)~ \left[1-g(t,p)^2\right]
\end{equation*}
and the second derivative is
\begin{align*}
	\partial_{t}^2 f(t,p) =&~ -\pi^2 ~ \sin\left(p~\pi~g(t,p)\right) ~ \left[1-g(t,p)^2\right]^2 \nonumber \\
	&\quad -2\frac{\pi}{p}~\cos\left(p~\pi~g(t,p)\right)~\left[g(t,p)-g(t,p)^3\right] \text{.}
\end{align*}
Again, we set $p=100$ and the time shift as above for the first and second derivative to yield 
\begin{equation*}
	\partial_{t} u(t) = \partial_{t} f(t-700,100) \quad \text{and} \quad \partial_{t}^2 u(t) = \partial_{t}^2 f(t-700,100) \text{.}
\end{equation*}

We visualize the two-sided sweep signal in Fig. \ref{fig:sweep_results} (a), and the resulting responses of three sub-circuits, namely voltages $z_{2}(t)$, $z_{6}(t)$ and $z_{10}(t)$, in Fig. \ref{fig:sweep_results} (b) to (d). We remark that the colors of input signals and their responses match. The response of ladder type 1, Fig. \ref{fig:sweep_results} (b), shows a typical low-pass filter behavior because the (high-frequency) oscillations are reduced from one sub-circuit to the next one. In Fig. \ref{fig:sweep_results} (c), we notice for $z_{2}$ a filtering behavior for high-frequency parts. Furthermore, we find vibrations are kicked in $z_{6}$ and $z_{10}$ when $\mu(t) = \partial_{t} u(t)$ transits from steady (zero) to oscillation and later again for the backward transition. We find a similar behavior for this ladder type in the first example, see Fig. \ref{fig:step_input_output_results} (c). Hence, such a dynamics seems to be an inherit element of this ladder type. In case of ladder type 3, the shape of input function $\mu(t) = \partial_{t}^2 u(t)$ occurs in the observed signals in Fig. \ref{fig:sweep_results} (d), too. Thus, the signal is put through the ladder only less with filtering. 

In a nutshell, we designed our three ladder types with the same electrical component values and we figured typical dynamics like charging and discharging, see Fig. \ref{fig:step_input_output_results} (b), and significant oscillations Fig. \ref{fig:step_input_output_results} (c) and Fig. \ref{fig:sweep_results} (c), (d). However, each ladder type exposes its own specific dynamical behavior and thus, they need to be analyzed in particular.

\section{Extension: Discrete Telegraph Line Equation}
\label{sec:telegraph_line}
The telegraph line equation\sidenote{It is noted as Telegrapher's equation in some texts, too.} and its corresponding circuit model, see Fig.  \ref{fig:telegraphers_ladder}, are common approaches to describe the voltage and current dynamics in an electrical transmission line, see the articles \cite{article:veigend2018model,article:shanak2020mathematical}. As transmission lines are spatially continuous, the original telegraph line equation is formulated as a partial differential equation (PDE). The circuit consists of a resistance $r$ and an inductance $\ell$ along the signal wire and a resistance $r_{s}$\sidenote{A conductance $G$ is often noted instead of our resistance $r_{s}$ with relation $G = \frac{1}{r_{s}}$.} and a capacitance $c_{n}$ between signal and return wire. In this section, we derive a state space formulation of the spatially discrete telegraphers model analog to the section \ref{sec:circuit_modeling}.

At the shunt we have a parallel connection with current $i_{s_1}$ and $i_{s_2}$ through the resistor and capacitor, respectively. Hence, we note the voltage 
\begin{align*}
	z_{n}(t) =&~ r_{s,n} i_{s_1,n} \quad \text{and} \\ 
	z_{n}(t) =&~ \frac{1}{c_{n}} \int_{0}^{t} i_{s_2,n}(\tau) d\tau 
\end{align*}
and we compute the total current in terms of the voltage and its derivative as 
\begin{equation*}
	i_{s,n}(t) =~  i_{s_1,n}(t) +  i_{s_2,n}(t) =~ \frac{z_{n}(t)}{r_{s,n}} + c_{n} \partial_{t} z_{n}(t) \text{.}
\end{equation*}
We have the voltage across the resistance $r$ and inductance $\ell$ as in Eq. \eqref{eq:rlc_cap_volt_r_l} and so we note
\begin{align}
	v_{n}(t) =&~  \left[r_{n} + \ell_{n}\partial_{t}\right] ~ \sum_{k=n}^{N} \left[ \frac{z_{k}(t)}{r_{s,k}} + c_{k} \partial_{t} z_{k}(t)\right] \nonumber \\
	=&~  \sum_{k=n}^{N}  \left[ \frac{r_{n}}{r_{s,k}} + \left(r_{n}~c_{k} + \frac{\ell_{n}}{r_{s,k}}\right) \partial_{t} +  \ell_{n} ~  c_{k} ~ \partial^2_{t} \right]   z_{k}(t) \text{.}
\end{align}
We insert the right-hand side of the previous equation in Eq. \eqref{eq:kirchhoff_voltage} in order to note the sum voltages in each sub-circuit as 
\begin{align}
	0 =~ \sum_{k=n}^{N}  &\left[ \ell_{n} ~  c_{k} ~ \partial^2_{t} z_{k}(t) +  \left(r_{n}~c_{k} + \frac{\ell_{n}}{r_{s,k}}\right) \partial_{t} z_{k}(t) +  \frac{r_{n}}{r_{s,k}} ~z_{k}(t)  \right]  \nonumber \\
	&\quad +  z_{n}(t) - z_{n-1}(t) \label{eq:telegraphers_shunt_ode_n}
\end{align}
and
\begin{equation*}
	u(t) = \sum_{k=1}^{N}  \left[ \ell_{1} ~  c_{k} ~ \partial^2_{t}  +  \left(r_{1}~c_{k} + \frac{\ell_{1}}{r_{s,k}}\right) \partial_{t} +  \frac{r_{n}}{r_{s,k}}   \right] ~z_{k}(t) +~  z_{1}(t) \text{.}
\end{equation*}
We see that Eq. \eqref{eq:telegraphers_shunt_ode_n} looks similar to Eq. \eqref{eq:rl-c_shunt_ode_n} of circuit type 1 but here we have additional terms for $z_{k}$ and $\partial_{t} z_{k}$. Hence, we note the second-order system \eqref{eq:rlc_second_order_1} as described in section \ref{sec:state_space_model_second_order} with triangular matrices
\begin{equation*}
	\mathbf{M} = \left[\ell_{n}~c_{k}\right] \quad \text{and} \quad \mathbf{D}=\left[r_{n}~c_{k} + \frac{\ell_{n}}{r_{s,k}}\right]
\end{equation*}
for $n\in \{1,\ldots,N\}$ and $k\in\{n,\ldots,N\}$. In case of the \textit{stiffness} matrix, we see that it consists of two parts
\begin{equation*}
	\mathbf{S} = \mathbf{F} + \mathbf{R} 	
\end{equation*}
with finite difference matrix $\mathbf{F}$ as in Eq. \eqref{eq:finite_diff_stencil} and a triangular matrix ${\mathbf{R}} = \left[\frac{r_{n}}{r_{s,k}} \right]$ containing the ratio between the resistance of the signal wire and the resistance between signal and return wire.

\begin{figure}[t!]
	\centering
	\begin{tikzpicture}[scale=1,line cap=round,line join=round,x=1.0cm,y=1.0cm]	
		\def\dxoff{7};
		\def\dyoff{3.8};
		
		
		\draw[line width=1.3, color=gray] (0,0) -- (0.5,0);
		\draw[line width=1.3, color=gray] (1.5,0) -- (2.0,0);
		\draw[line width=1.3, color=gray] (2.9,0) -- (6.8,0);
		
		\draw[line width=1.3, color=gray] (4.5,0) -- (4.5,-0.9);
		\draw[line width=1.3, color=gray] (4.5,-1.9) -- (4.5,-2.6);
		
		\draw[line width=1.3, color=gray] (5.7,0) -- (5.7,-1.3);
		\draw[line width=1.3, color=gray] (5.7,-1.5) -- (5.7,-2.6);

		\draw[line width=1.3, color=gray] (0,-2.6) -- (6.8,-2.6);

		\draw [line width=1.0] (-0.1, 0) circle [radius = 0.1];
		\draw [line width=1.0] ( 6.9, 0) circle [radius = 0.1];
		\draw [line width=1.0] (-0.1,-2.6) circle [radius = 0.1];
		\draw [line width=1.0] ( 6.9,-2.6) circle [radius = 0.1];
		
		\draw[line width=2, color=blue,rounded corners] (0.5,-0.2) rectangle (1.5,0.2);

		\draw[line width=2, color=blue] (2.0,0.) arc ( 180:0:0.15);
		\draw[line width=2, color=blue] (2.3,0.) arc ( 180:0:0.15);
		\draw[line width=2, color=blue] (2.6,0.) arc ( 180:0:0.15);
		
		\draw[line width=2, color=blue,rounded corners] (4.3,-0.9) rectangle (4.7,-1.9);
		
		\draw[line width=2, color=blue]  (5.4,-1.3) -- (6,-1.3);
		\draw[line width=2, color=blue]  (5.4,-1.5) -- (6,-1.5);

		\node at ( 1., -0.55) {$r_{n}$};
		\node at ( 2.45, -0.4) {$\ell_{n}$};
		\node at ( 3.7, -1.7) {$r_{s,n}$};
		\node at ( 5.2, -1.7) {$c_{n}$};

		\draw [->, line width=1.3, color=magenta] (3.1,0.05) -- (3.9,0.05);
		\draw [->, line width=1.3, color=magenta] (5.8,0.05) -- (6.6,0.05);
		\draw [->, line width=1.3, color=magenta] (4.42,-0.1) -- (4.42,-0.8);
		\draw [->, line width=1.3, color=magenta] (5.62,-0.1) -- (5.62,-0.8);
		
		\node at (3.5, 0.3) {$i_{m,n}$};
		\node at (6.2, 0.3) {$i_{m,n}$};
		\node at (4.0, -0.4) {$i_{s_1,n}$};
		\node at (5.2, -0.4) {$i_{s_2,n}$};

		\draw [->, line width=1.3, color=olive]	(-0.1, -0.2) -- (-0.1, -2.4);
		\draw [->, line width=1.3, color=olive]	( 6.9, -0.2) -- ( 6.9, -2.4);
		\draw [->, line width=1.3, color=olive]	( 0.45, 0.5) -- ( 2.95, 0.5);
		\node at (-0.6, -1.3) {$z_{n-1}$};
		\node at ( 7.3, -1.3)  {$z_{n}$};
		\node at ( 1.6, 0.7) {$v_{n}$};
		
	\end{tikzpicture}
	\caption{Circuit of a spatially discrete Telegrapher's or transmission line model.}
	\label{fig:telegraphers_ladder}
\end{figure}

In the next step, we compose the state space equations \eqref{eq:state_space} in accordance with Eq. (\ref{eq:rlc_second_order_2}, \ref{eq:first_order_submatrices}). We note the submatrices as 
\begin{align*}
	\mathbf{A_{I}} =&~ -\mathbf{M}^{-1} \mathbf{S}  = \underbrace{\left[-\mathbf{M}^{-1} ~  \mathbf{F}\right]}_{\text{matrix shape } \mathbf{L}} +  \underbrace{\left[-\mathbf{M}^{-1} ~ \mathbf{R}\right]}_{\text{matix shape } \mathbf{P}} \quad \text{and} \\[1ex]
	\mathbf{A_{II}} =&~ -\mathbf{M}^{-1} \mathbf{D} = \underbrace{\left[ -\mathbf{M}^{-1} ~ \mathbf{D}_{1}\right]}_{\text{matix shape } \mathbf{P}} + \underbrace{\left[-\mathbf{M}^{-1} ~ \mathbf{D}_{2}\right]}_{\text{matix shape } \mathbf{P}}  \nonumber
\end{align*}
with $\mathbf{D}_{1}:=\left[r_{n}~c_{k}\right]$ and $\mathbf{D}_{2}:=\left[\frac{\ell_{n}}{r_{s,k}}\right]$; and we see $\mathbf{A_{I}}$ consists of two different matrix shapes, $\mathbf{L}$ and $\mathbf{P}$, while  $\mathbf{A_{II}}$ has the shape $\mathbf{P}$. In case of uniform component values, meaning
\begin{equation*}
	r_{i} = r_{j} ~\text{,}~ r_{s,i} = r_{s,j} ~\text{,} ~ \ell_{i} = \ell_{j} ~\text{and}~ c_{i} = c_{j} ~\text{for}~ i,j \in \{1,\ldots,N\} ~\text{,}
\end{equation*}
we formulate the right-hand side of $\dot{x}_{2}$ as 
\begin{fullwidth}
	\begin{subequations}
		\begin{align}
			\dot{x}_{1}(t) =&~ x_{2}(t) \quad \text{,} \\
			\dot{x}_{2}(t) =&~ \underbrace{\left(\frac{1}{\ell c}\mathbf{\tilde{L}} - \frac{r}{r_{s}~\ell~c} \mathbf{I}\right)}_{=\mathbf{A_{I}}} ~ x_{1}(t)  
			\underbrace{-\left(\frac{r}{\ell} + \frac{1}{r_{s}~c} \right) ~ \mathbf{I}}_{=\mathbf{A_{II}}} ~ x_{2}(t) + 
			\frac{1}{\ell~c} ~
			\begin{pmatrix}
				1\\
				0\\
				\vdots \\
				0
			\end{pmatrix}
			\mu(t) \quad \text{.}
		\end{align}		
		\label{eq:state_space_telegrapher}
	\end{subequations}
\end{fullwidth}

The telegraph line equation is formulated as a second-order PDE\sidenote{In the literature we often find first-order formulations including the current and voltage, see \cite{article:veigend2018model}. Here we note it in a compact form only with the voltage dynamics, which is derived from such a current and voltage formulation.} 
\begin{equation*}
	\ell~c \frac{\partial^2}{\partial t^2} w(t,x) + \left(r~c + \frac{\ell}{r_{s}}\right)\frac{\partial}{\partial t} w(t,x) - \frac{\partial^2}{\partial x^2} w(t,x) + \frac{r}{r_{s}} w(t,x) = 0
\end{equation*}
with voltage $w \in (0,\chi) \times (0,T)$ and $\chi, T > 0$ and equipped with suitable initial and boundary conditions. Our second-order ordinary differential equation with uniform component values 
\begin{equation}
	\ell c ~ \ddot{z}(t) + \left(r~c + \frac{\ell}{r_{s}}\right) ~ \mathbf{I} ~ \dot{z}(t) + \left(\frac{r}{r_{s}} \mathbf{I} - \mathbf{\tilde{L}} \right) ~ z(t) = 
	\begin{pmatrix}
		1\\
		0\\
		\vdots \\
		0
	\end{pmatrix}
	\mu(t)
	\label{eq:telegraph_second_order}
\end{equation}
represents a spatially discrete version of the telegraph line PDE. As a side note, we mention that the right-hand side of Eq. \eqref{eq:telegraph_second_order} with the input signal $\mu(t)$ would correspond to a Dirichlet boundary condition for the PDE model as $w(t,0) = \mu(t)$.


\section*{Conclusion and Outlook}

In this article, we describe a state space modeling for three types of electrical ladders. Here, we formulate the system matrix $\mathbf{A}$ and its sub-matrices $\mathbf{A}_{I}$ and $\mathbf{A}_{II}$ for each ladder type and, in particular, for non-uniform coefficients. We present two simulations for our ladder models with step and oscillation input signals to emphasize the dynamical behavior of each ladder type.

As this article is intended to focus only on the state space modeling, various related topics are missing. For example, we do not provide here an (in-depth) analysis of the proposed models, i.e. eigenvalue and eigenvector analysis. In the sense of control systems, we need to discuss popular matrix-based methods for controllability, observability, system identification and feed-forward and feedback control design. For these points, we need check how to treat our matrix shapes of $\mathbf{A}_{I}$ and $\mathbf{A}_{II}$ in a smart way to simplify such computations. We also find interesting study cases in the field of scientific machine learning to compute the system models, see again article \cite{article:peters2023dynamic}. Moreover, the story of circuit configurations is not complete yet with three ladder types, i.e. we skipped parallel connections in the main branch. Hence, we find several open points need to be discussed in latter contributions.


\section*{Source Code}
The simulations in Section \ref{sec:simulation_examples} are implemented with \textsc{Julia} programming \cite{article:bezanson2017julia}. As part of this research project, the author develops the Julia package \textit{Ellasy.jl} 
\begin{center}
	 \url{https://github.com/stephans3/Ellasy.jl} 
\end{center}
to build the state space models of electrical ladder systems. The source code of our examples is stored in \textit{EllasyExamples.jl}
\begin{center}
	\url{https://github.com/stephans3/EllasyExamples.jl}.
\end{center}
As part of our simulations, we use the libraries \textit{DifferentialEquations.jl} \cite{software:rackauckas2020sciml} to solve the ODE numerically with \textit{Tsit5} solver \cite{article:tsitouras2011runge}, and \textit{Makie.jl} \cite{article:danisch2021makie} to create the figures \ref{fig:step_input_output_results} and \ref{fig:sweep_results}.


\section*{Acknowledgments}	
This article is inspired by student projects and theses at the University of Applied Sciences Ravensburg-Weingarten, which were supervised by Lothar Berger and S. Scholz.




\end{document}